\documentclass[11pt]{article}
\pdfoutput=1
\usepackage{jheppub}

\usepackage{url}

\allowdisplaybreaks[2]

\def\cA{\mathcal{A}}
\def\cB{\mathcal{B}}
\def\cC{\mathcal{C}}
\def\cD{\mathcal{D}}

\def\cJ{\mathcal{J}}

\def\cN{\mathcal{N}}
\def\cO{\mathcal{O}}

\def\cW{\mathcal{W}}

\def\mint{\int_{-\infty}^\infty\!\cdots\!\int_{-\infty}^\infty}

\newcommand{\be}{\begin{equation}}
\newcommand{\ee}{\end{equation}}
\newcommand{\ba}{\begin{aligned}}
\newcommand{\ea}{\end{aligned}}

\def\vev#1{\langle #1 \rangle}

\def\bra#1{\left\langle #1 \right|}
\def\ket#1{\left| #1 \right\rangle}

\DeclareMathOperator{\Ai}{Ai}

\def\({\left(}
\def\){\right)}

\newcommand{\pd}{\partial}

\DeclareMathOperator{\Tr}{Tr}

\DeclareMathOperator{\Pexp}{Pexp}

\newcommand{\re}{{\rm e}}
\newcommand{\ri}{{\rm i}}
\newcommand{\rd}{{\rm d}}

\def\cob{\delta}

\newcommand{\hf}{\frac{1}{2}}

\def\til#1{\widetilde{#1}}

\def\bra{\langle}
\def\ket{\rangle}

\def\lrya{\leftrightarrow}

\def\ka{\kappa}
\def\h#1{\widehat{#1}}

\def\rt#1{\sqrt{#1}}
\def\sitarel#1#2{\mathrel{\mathop{\kern0pt #1}\limits_{#2}}}

\newdimen\tableauside\tableauside=1.0ex
\newdimen\tableaurule\tableaurule=0.4pt
\newdimen\tableaustep
\def\phantomhrule#1{\hbox{\vbox to0pt{\hrule height\tableaurule width#1\vss}}}
\def\phantomvrule#1{\vbox{\hbox to0pt{\vrule width\tableaurule height#1\hss}}}
\def\sqr{\vbox{%
  \phantomhrule\tableaustep
  \hbox{\phantomvrule\tableaustep\kern\tableaustep\phantomvrule\tableaustep}%
  \hbox{\vbox{\phantomhrule\tableauside}\kern-\tableaurule}}}
\def\squares#1{\hbox{\count0=#1\noindent\loop\sqr
  \advance\count0 by-1 \ifnum\count0>0\repeat}}
\def\tableau#1{\vcenter{\offinterlineskip
  \tableaustep=\tableauside\advance\tableaustep by-\tableaurule
  \kern\normallineskip\hbox
    {\kern\normallineskip\vbox
      {\gettableau#1 0 }%
     \kern\normallineskip\kern\tableaurule}%
  \kern\normallineskip\kern\tableaurule}}
\def\gettableau#1{\ifnum#1=0\let\next=\null\else
\squares{#1}\let\next=\gettableau\fi\next}

\title{Exact results for ABJ Wilson loops and open-closed duality}

\author[a]{Yasuyuki Hatsuda}
\author[b]{and Kazumi Okuyama}

\affiliation[a]{D\'epartement de Physique Th\'eorique et Section de Math\'ematiques\\
Universit\'e de Gen\`eve, Gen\`eve, CH-1211 Switzerland}
\affiliation[b]{Department of Physics, \\
Shinshu University, Matsumoto 390-8621, Japan}

\emailAdd{yasuyuki.hatsuda@unige.ch} 
\emailAdd{kazumi@azusa.shinshu-u.ac.jp}

\abstract{
We find new exact relations between the partition function and vacuum expectation values (VEVs)
of 1/2 BPS Wilson loops in ABJ theory, which
allow us to predict the large $N$ expansions of the 1/2 BPS Wilson loops
from known results of the partition function.
These relations are interpreted as an open-closed duality
where the closed string background is shifted by the insertion of Wilson loops due to a back-reaction.
Using the connection between ABJ theory
and the topological string on local $\mathbb{P}^1\times \mathbb{P}^1$,
we explicitly write down non-trivial relations between open and closed string amplitudes. 
}

\begin{document}

\maketitle

\renewcommand{\thefootnote}{\arabic{footnote}}
\setcounter{footnote}{0}
\setcounter{section}{0}

\section{Introduction}\label{sec:intro}
This work addresses exact computations of Wilson loops in ABJ(M) theory \cite{ABJM, ABJ}.
Our analysis heavily relies on the supersymmetric localization \cite{Pestun}.
The ABJ theory is a 3d $\cN=6$ superconformal Chern-Simons-matter theory with quiver
gauge group $U(N_1)_k \times U(N_2)_{-k}$,
which  describes a low energy effective theory on multiple M2-branes.
When the ranks of the two gauge groups are equal,
the theory is especially called the ABJM theory.
As shown in \cite{KWY1, Jafferis, HHL1}, 
the partition function of ABJ theory (and more generally
for a wider class of Chern-Simons-matter theories) 
on a three-sphere reduces to a matrix integral by the localization.
A basic problem is then to understand the large $N$ behavior of the obtained matrix model.

To extract the information at large $N$ from the ABJ matrix model,
one can use two remarkable facts.
One is a connection with the topological string on a particular Calabi-Yau three-fold, known as 
local  $\mathbb{P}^1 \times \mathbb{P}^1$.
This connection is a consequence of a chain of dualities:
The ABJ matrix model is related to a matrix model on a lens space $L(2,1)$
by analytic continuation \cite{MP-top}.
It is known that this lens space matrix model is dual to the topological string 
on local $\mathbb{P}^1 \times \mathbb{P}^1$ at large $N$ \cite{AKMV}.
As a result, the large $N$ expansion in the ABJ matrix model is captured by the topological string on local $\mathbb{P}^1 \times \mathbb{P}^1$.
The all-genus free energy in the 't Hooft limit was indeed computed in \cite{DMP1} 
by using the topological string technique.
The planar free energy shows the  $N^{3/2}$ behavior,
which correctly reproduces the expected number of degrees of freedom of $N$ M2-branes.
The same behavior was also confirmed by the direct saddle-point analysis in the M-theory limit \cite{HKPT}.

Another key fact is an unexpected relation to a non-interacting quantum Fermi-gas system.
It was shown in \cite{MP} that the ABJM partition function is regarded as the partition function
of an ideal quantum Fermi-gas with an unconventional Hamiltonian.
This  Fermi-gas picture allows us to analyze the system by
the technique of  statistical mechanics.
The important point is that the role of the Planck constant in this quantum system is played by
the Chern-Simons level, and the semi-classical limit corresponds to the strong coupling
limit in ABJM theory.
Therefore we can get the strong coupling results from the semi-classical analysis in the Fermi-gas system.
Putting the various pieces of information together (see \cite{HMO-review} for a review and references therein),
it was finally shown in \cite{HMMO} that the complete large $N$ expansion, including both 
worldsheet instanton and membrane instanton corrections, of the partition function 
are determined 
by the (refined) topological string on local $\mathbb{P}^1 \times \mathbb{P}^1$ in a highly non-trivial way.
Quite interestingly, there is a pole cancellation mechanism between the worldsheet instanton corrections
and the membrane instanton corrections \cite{HMO2}, which guarantees the theory to be well-defined for any
value of $k$.
The Fermi-gas approach was also extended to the ABJ matrix model \cite{MaMo, HoOk} (see also \cite{AHS, Honda1}), 
and it was revealed that the large $N$ expansion is again determined by
the topological string on local $\mathbb{P}^1\times \mathbb{P}^1$.

In this paper, we study  circular BPS Wilson loops in ABJ theory on $S^3$.
There are two kinds of BPS Wilson loops in ABJ theory.
One preserves 1/6 of supersymmetries \cite{DPY, CW, RSY}, while the other a half of them \cite{DT}. 
We here focus on the latter since it has a much simpler structure than the former.
The vacuum expectation values (VEVs) of BPS Wilson loops
can be exactly computed by the localization.
The Fermi-gas approach for the Wilson loops in ABJM theory 
was first proposed in \cite{KMSS} and 
a similar formalism was further developed in \cite{HHMO}
especially for the 1/2 BPS Wilson loops.
It was shown in \cite{MP-top} that the large $N$ expansions of the 1/2 BPS Wilson loops
are explained by the \textit{open} topological string on local $\mathbb{P}^1 \times \mathbb{P}^1$.
The Seiberg-like duality of the ABJ Wilson loops was discussed in \cite{HNS}.

It is known that the ABJ partition function has several good properties.
For some particular values of $k$, the generating function of the partition function
can be written in closed form \cite{CGM, GHM2}.
This fact enables us to predict the exact values of the partition function without performing
the matrix integral.
It was also found in \cite{GHM2} that the generating function of the ABJ partition function satisfies
beautiful functional relations. 
Therefore it is natural to ask whether the Wilson loops in ABJ theory also have some nice properties or not.
As we will show in this paper, the answer is yes: We find remarkable exact relations among the 
1/2 BPS Wilson loops!

\begin{figure}[tb]
\begin{center}
\resizebox{110mm}{!}{\includegraphics{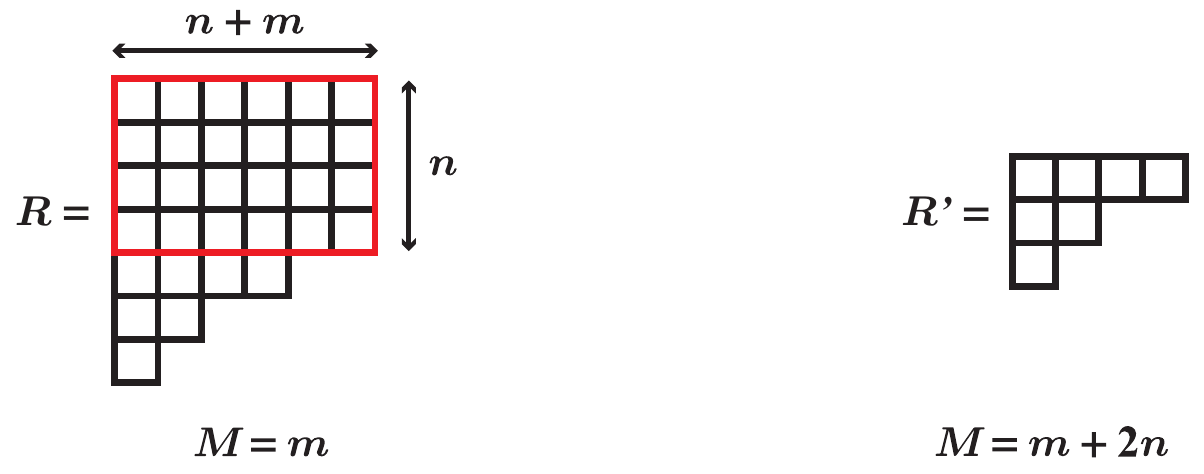}}
\end{center}
\vspace{-0.5cm}
  \caption{The VEV of the 1/2 BPS Wilson loop with the left Young diagram $R$
			in $M=m$ ABJ theory is related to the one with the right diagram $R'$ in
			$M=m+2n$ ABJ theory. In other words, the effect of the red rectangular part in $R$
			results in adding the $2n$ fractional M2-branes after the back-reaction.
			The relation \eqref{eq:exact-WL} corresponds to the simplest case that 
			$n=1$, $m=0$ and $R'$ is the trivial representation.}
  \label{fig:Young}
\end{figure}

Based on the previous analysis in \cite{MaMo, HoOk, HHMO}, we here find new exact relations between
the partition function and the VEVs of the 1/2 BPS Wilson loops.
In the simplest case, the VEV of the 1/2 BPS Wilson loop with the fundamental representation 
in $U(N)_k \times U(N)_{-k}$ ABJM theory
is \textit{exactly} related to the partition function of $U(N-1)_k \times U(N+1)_{-k}$ ABJ theory.
Namely, the normalized VEV of the fundamental ABJM Wilson loop is given by
\be
\ba
\vev{W_{\tableau{1}}^{(1/2)}}_{N,k}^\text{ABJM}
=\frac{1}{2\sin \frac{2\pi}{k}} \biggl| \frac{Z^\text{ABJ} (N-1,N+1,k)}{Z^\text{ABJM} (N,k)} \biggr|.
\ea
\label{eq:exact-WL} 
\ee
We conjecture that this relation holds for \textit{any} $N$ and $k(>2)$.
In the large $N$ limit, the all-order perturbative $1/N$ expansion of the free energy can be resummed,
and it results in the Airy functional form \cite{FHM}.
As shown in \cite{KMSS}, the VEVs of the 1/2 BPS Wilson loops are also resummed as
the Airy function, and the final result in \cite{KMSS} is consistent with our exact relation \eqref{eq:exact-WL}.
We stress that our relation \eqref{eq:exact-WL}, however, contains all the non-perturbative corrections,
and it is true even for finite $N$.
As we will see in section~\ref{sec:exact-rel}, the relation \eqref{eq:exact-WL} is only a tip of the iceberg.
We find many similar relations for higher representations in the ABJ Wilson loops, as shown in Fig.~\ref{fig:Young}.
We also find a determinant formula that computes the VEVs of the 1/2 BPS Wilson loops
with general representations only from ``hook'' representations.\footnote{While  preparing the draft of this paper,
we were informed by Sanefumi Moriyama 
that the determinant formula 
 can be proved for general ABJ theory \cite{Mto}. We would like to
thank him for telling us before submission of their paper.}
This is a natural generalization of the result in \cite{HHMO} to the ABJ Wilson loops.

The relation \eqref{eq:exact-WL} has the following interpretation:
From the viewpoint of type IIA string theory
on $AdS_4\times \mathbb{CP}^3$, which is holographically dual to ABJ
theory on $S^3$,
the left hand side of \eqref{eq:exact-WL} is related to the fundamental open string.
On the other hand, the right hand side  of \eqref{eq:exact-WL}
is purely explained by
a change of the closed string background,
where 
the value of
the NSNS B-field flux
through a two-cycle $\mathbb{CP}^1\subset\mathbb{CP}^3$
\begin{align}
 \int_{\mathbb{CP}^1}B_\text{NS}=\hf-\frac{M}{k}, \qquad M=N_2-N_1.
\end{align}
is shifted from $M=0$ to $M=2$ by inserting the fundamental Wilson loop.
In this sense, we can regard \eqref{eq:exact-WL} as a kind of \textit{open-closed dualities}.
In the brane setup, $M$ counts the number of fractional M2-branes \cite{ABJ}.
Our relation \eqref{eq:exact-WL} implies that the effect of the brane that describes the fundamental
Wilson loop in ABJM theory results in two fractional branes in ABJ theory due to the back-reaction.%
\footnote{A similar picture has been noted in \cite{MaMo}. It was shown there that the ABJ partition function
is related to quantities similar to Wilson loops in ABJM theory. We would like to emphasize that  ABJ(M) Wilson loops 
themselves are also related to the ABJ(M) partition function.}
Roughly speaking, if the dimension of the representation in the Wilson loop grows, then the number of the fractional branes
also increases, as in Fig.~\ref{fig:Young}.%
\footnote{However, the increase of the fractional branes is somewhat obscure because ABJ theory has Seiberg-like duality, 
which relates $U(N)_k \times U(N+M)_{-k}$ theory to $U(N+k-M)_k \times U(N)_{-k}$ theory.}

The open-closed duality here is also understood from the topological string perspective.
As mentioned above, the left hand side of
\eqref{eq:exact-WL} corresponds to the open topological string \cite{MP-top, GKM, HHMO},
while the right hand side is captured by the closed topological string \cite{HMMO, HoOk}.
Thus, we find that the open topological string amplitude on local $\mathbb{P}^1 \times \mathbb{P}^1$
is related to the closed string amplitude on the same Calabi-Yau.
As explained in \cite{ADKMV},
the effect of the open string generally
leads to a shift of certain moduli of the closed string, \textit{i.e.},
the open string amplitude is schematically written as 
\begin{align}
 Z_\text{open}(Q)=\frac{Z_\text{closed}(Q')}{Z_\text{closed}(Q)},
\label{eq:open-closed}
\end{align}
where $Q$ and $Q'$ are moduli of the (same) Calabi-Yau manifold.%
\footnote{In the open-closed duality, the open (+ closed) string theory on a Calabi-Yau manifold $X$
is, in general, mapped to the closed string theory on a \textit{different} CY $X'$.
Our result states that if $X$ is local $\mathbb{P}^1 \times \mathbb{P}^1$, the geometry $X'$
appearing after integrating out some particular open string sector 
is also local $\mathbb{P}^1 \times \mathbb{P}^1$, 
but its moduli is different from those of $X$.}
Our relation is a concrete realization
of this open-closed duality at the quantitative level.
In section~\ref{sec:open-closed}, we will explicitly show that \eqref{eq:exact-WL} leads to the non-trivial relation
\eqref{eq:open-closed-2} between the open and closed string amplitudes.
Also, the relation in Fig.~\ref{fig:Young} is easily translated into the topological string language \eqref{eq:open-open}.

The organization of this paper is as follows.
In the next section, we briefly review the 1/2 BPS Wilson loops in ABJ theory.
We mainly use the Fermi-gas formalism proposed in \cite{MaMo}.
In section~\ref{sec:exact-rel}, we demonstrate that the ABJ Wilson loops for different $M$'s are interrelated.
In some cases, the relations reduce to the ones between the partition function and the Wilson loops,
and we can interpret them as an open-closed duality.
In section~\ref{sec:special}, we show additional results for $k \in 4\mathbb{N}$.
In section~\ref{sec:open-closed}, we consider a consequence of the relations found in section~\ref{sec:exact-rel}.
We explicitly show that the open topological string partition function is related to
the closed one.
We conclude in section~\ref{sec:conclusion} 
and comment on some future directions.

\section{A review of ABJ Wilson loops}\label{sec:review}
Let us start with a review of the Wilson loops in ABJ theory.

\subsection{ABJ Wilson loops}
In ABJ theory, two kinds of circular BPS Wilson loops are widely studied in the literature.
One preserves only 1/6 of supersymmetries, while the other a half of supersymmetries. 
The 1/6 BPS Wilson loops is explicitly given by
\be
\ba
W_R^{(1/6)}=\Tr_R \Pexp \left[ \int \! \rd s \( \ri A_\mu \dot{x}^\mu+\frac{2\pi}{k} |\dot{x}| M^I_J C_I \bar{C}^J \) \right],
\ea
\ee
where $x^\mu(s)$ parametrizes a great circle of $S^3$, and $C_I$ ($I=1,2,3,4$) are scalar fields in the four bi-fundamental chiral multiplets. 
The matrix $M^I_J$ is chosen in order to preserve the supersymmetry.
The construction of the 1/2 BPS Wilson loops is more complicated. See \cite{DT} for detail. 

The localization method allows us to reduce the path integral to a finite-dimensional matrix integral.
The partition function of ABJ theory on $S^3$ is exactly given by
\be
\ba
Z^\text{ABJ}(N_1,N_2,k)=\frac{\ri^{-\frac{1}{2}(N_1^2-N_2^2)}}{N_1! N_2!}
\int \! \prod_{i=1}^{N_1} \frac{\rd \mu_i}{2\pi} \prod_{j=1}^{N_2} \frac{\rd \nu_j}{2\pi} \, \re^{-\frac{\ri k}{4\pi} (\sum_i \mu_i^2-\sum_{j} \nu_j^2)}\\
\times \frac{\prod_{i<j} (2\sinh \frac{\mu_i-\mu_j}{2})^2 (2 \sinh \frac{\nu_i-\nu_j}{2} )^2}
{\prod_{i,j} (2\cosh \frac{\mu_i-\nu_j}{2} )^2}.
\ea
\label{eq:Z-ABJ}
\ee 
In the analysis below, we always assume that $k>0$ and $N_1 \leq N_2$
without loss of generality.
Sometimes it is convenient to parametrize $N_1$ and $N_2$ by 
\be
N_1=N,\qquad N_2=N+M,  \qquad M \geq 0.
\ee
Physically, $M$ corresponds to the number of fractional M2-branes \cite{ABJ}.
It was shown in \cite{KWY1} that the VEVs of the 1/6 BPS Wilson loops are given by the insertion of 
an operator in the above matrix model
\be
\vev{W_R^{(1/6)}}_{N_1,N_2,k}=\vev{s_R(\re^{\mu_1},\dots, \re^{\mu_{N_1}})},
\ee
where $s_R(\re^{\mu_1},\dots, \re^{\mu_{N_1}})$ is the Schur polynomial with representation $R$ in $U(N_1)$.
The expectation values on the right hand side means the \textit{unnormalized} VEV for the ABJ matrix model \eqref{eq:Z-ABJ}.
Of course one can consider an insertion in the other gauge group $U(N_2)$.

The VEVs of the 1/2 BPS Wilson loops are also given by the insertion of the character of the supergroup $U(N_1|N_2)$ \cite{DT}.
Since the character of the supergroup $U(N_1|N_2)$ is given by the supersymmetric Schur polynomial,
we have
\be
\vev{W_R^{(1/2)}}_{N_1,N_2,k}=\vev{s_R(\re^{\mu_1},\dots, \re^{\mu_{N_1}}/\re^{\nu_1},\dots, \re^{\nu_{N_2}})},
\label{eq:half-vev}
\ee
where $s_R(\re^{\mu_1},\dots, \re^{\mu_{N_1}}/\re^{\nu_1},\dots, \re^{\nu_{N_2}})$ is the super Schur polynomial
associated with the representation $R$ in supergroup $U(N_1|N_2)$, which
is related to the standard Schur polynomial by
\be
s_\lambda(x/y)=\sum_{\mu, \nu} N_{\mu \nu}^\lambda s_\mu(x) s_{\nu^T}(y).
\ee
Here $N_{\mu \nu}^\lambda$ is the Littlewood-Richardson coefficient,
and $\lambda^T$ means the conjugate (or transposed) partition of $\lambda$.
Note that the super Schur polynomial satisfies  a conjugation formula
\be
s_\lambda(x/y)=s_{\lambda^T}(y/x).
\ee 

Before reviewing the exact computation of the VEVs of the 1/2 BPS Wilson loops,
let us explain a convention of representations.
In this paper, we often use the so-called Frobenius notation of representations.
The standard Frobenius notation for the partition $\lambda=[\lambda_1\lambda_2\lambda_3\dots]$
is denoted by $(a_1 \dots a_r|l_1 \dots l_r)$, where $a_q$ and $l_p$ are given by
\be
a_q=\lambda_q-q,\qquad l_p=\lambda^T_p-p.
\ee
Here $\lambda^T=[\lambda^T_1 \lambda^T_2 \lambda^T_3 \dots]$ is the conjugate partition of $\lambda$.
The maximal value $r$ is defined by
\be
r=\max \{s | \lambda_s-s \geq 0\}.
\ee
In \cite{MaMo}, a modification of the Frobenius notation was also introduced.
For a given non-negative integer $M$, we define $\widehat{a}_q$ and $\widehat{l}_p$ by
\be
\widehat{a}_q=\lambda_q-q-M,\qquad \widehat{l}_p=\lambda^T_p-p+M.
\label{eq:mFrobenius}
\ee
Then, the modified Frobenius notation of $\lambda$ is denoted by $(\widehat{a}_1\dots \widehat{a}_{r_M}|
\widehat{l}_1 \dots \widehat{l}_{r_M+M})$, where $r_M$ is now
\be
r_M=\max \{s | \lambda_s-s-M \geq 0\}.
\ee
If $\lambda_1 \leq M$, we define $r_M=0$, and denote the modified Frobenius notation by $(|\widehat{l}_1\dots \widehat{l}_M)$. 
Of course, for $M=0$, the modified Frobenius notation is identical to the standard Frobenius notation.
Let us see an example.
For the representation $R=[2,2,1]=\tableau{2 2 1}$, the standard Frobenius notation is $(10|20)$.
For $M=1$, the modified Frobenius notation of the same representation is $(0|31)$,
and for $M=2,3$, one finds $(|42)$ and $(|530)$, respectively.
See Fig.~1 in \cite{MaMo} for more details.

\subsection{Fermi-gas formalism}\label{sec:Fermi-gas}
It is not easy to evaluate the ABJ matrix model \eqref{eq:Z-ABJ} directly.
Fortunately, there is a powerful method to evaluate it, known as the Fermi-gas formalism \cite{MP}.
This formalism can be generalized to the Wilson loops \cite{KMSS, HHMO, MaMo}.
Here we will briefly review that formalism.

\subsubsection{Generating function with phase factor}
Our main goal is to evaluate the unnormalized VEVs \eqref{eq:half-vev} systematically.
To do so, we use the nice formalism in \cite{MaMo}.
Let us define a generating function of the VEVs of the $1/2$ BPS Wilson loops with representation $R$ by
\be
\cW_R^\text{phase}(\kappa, k, M)=\sum_{N=0}^\infty \kappa^{N} \vev{W_R^{(1/2)}}_{N_1=N, N_2=N+M, k}\,,
\label{eq:W_R-GC}
\ee
where we call $\kappa$ the fugacity by analogy with the grand canonical ensemble.
Below, we use both the fugacity $\kappa$ and the chemical potential $\mu=\log \kappa$ interchangeably.
We also introduce the grand canonical partition function by
\be
\Xi^{\text{phase}}(\kappa, k , M)=\sum_{N=0}^\infty \kappa^N Z^\text{ABJ}(N,N+M,k).
\label{eq:Xiphase}
\ee
In general,
the partition function $Z^\text{ABJ}(N,N+M,k)$ and the VEV $\vev{W_R^{(1/2)}}_{N, N+M, k}$ are 
complex-valued, and have a non-trivial phase. Hence we put the superscript ``phase'' in 
\eqref{eq:W_R-GC} and \eqref{eq:Xiphase}.
As shown in \cite{AHS,HoOk}, the ABJ partition function \eqref{eq:Z-ABJ} can be written as
\be
Z^\text{ABJ}(N,N+M,k)=\re^{\ri \theta_0(k,M)}\re^{\frac{\pi\ri }{2}MN} 
Z^\text{CS}(k,M)\h{Z}(N,N+M,k),
\ee
where $\re^{\ri \theta_0(k,M)}$ is a phase factor, 
\begin{align}
 \begin{aligned}
  \re^{\ri \theta_0(M,k)}=\exp\left[-\frac{\pi\ri M(M^2-1)}{6k}
+\frac{\pi\ri M(M-2)}{4}\right],
 \end{aligned}
\label{eq:th0}
\end{align}
and $Z^\text{CS}(k, M)$ 
is the partition function 
of the pure $U(M)$ Chern-Simons (CS) theory
\be
Z^\text{CS}(k,M)=k^{-M/2} \prod_{s=1}^{M-1} \( 2\sin \frac{\pi s}{k} \)^{M-s}.
\ee 
Note that $\h{Z}(N,N+M,k)$ always takes a real value and obeys
\begin{align}
 \h{Z}(0,M,k)=1.
\end{align}
It is also convenient to introduce
generating functions of the absolute value $|Z^\text{ABJ}(N,N+M,k)|$
and 
of the rescaled partition function divided by the pure CS factor,
\be
\ba
\Xi_0(\kappa,k,M) &=\sum_{N=0}^\infty\kappa^N |Z^\text{ABJ}(N,N+M,k)|
,\\
\Xi(\kappa,k,M) &=\frac{\Xi_0(\kappa,k,M)}{Z^\text{CS}(k,M)}=1+\sum_{N=1}^\infty 
\kappa^N\h{Z}(N,N+M,k),
\label{eq:Xi0}
\ea
\ee
but for the moment we will consider the grand partition function 
\eqref{eq:Xiphase} with phase.
For the ABJM case $M=0$, the phase factor and the pure CS partition function
are trivial,
and all the definitions in \eqref{eq:Xiphase} and
 \eqref{eq:Xi0} are identical
\be
\Xi^\text{phase}(\kappa,k,0)=\Xi_0(\kappa,k,0)=\Xi(\kappa,k,0).
\ee
However, they are different for the general ABJ case with $M\not=0$.

\paragraph{Formalism of Matsumoto-Moriyama.}
In \cite{MaMo}, it was shown that the generating function \eqref{eq:W_R-GC} is given by
a determinant of an $(M+r_M) \times (M+r_M)$ matrix,
\be
\cW_R^\text{phase}(\kappa, k, M)=\Xi_0(\kappa, k, 0)
\det\Bigl((H_{\h{l}_p,M-q})_{1\leq q\leq M}\Bigl|(\til{H}_{\h{l}_p,\h{a}_q})_{1\leq q\leq r_M}\Bigr)_{1\leq p\leq M+r_M}.
\label{eq:W_R-GC-det}
\ee
In particular, the grand partition function is given by
\be
\Xi^\text{phase}(\kappa, k, M)=\Xi_0(\kappa, k, 0)
\det\Bigl(H_{M-p,M-q} \Bigr)_{1\leq p,  q\leq M}.
\label{eq:Xiphase-det}
\ee
Here the matrix elements $H_{m,n}$ and $\til{H}_{m,n}$ are  given by
\be
\ba
H_{m,n}&=E_{m+\hf}(\nu) \circ \frac{1}{1+\kappa Q(\nu, \mu) \circ P(\mu,\nu')} \circ E_{-n-\hf}(\nu'), \\
\til{H}_{m,n}&=\kappa E_{m+\hf}(\nu) \circ \frac{1}{1+\kappa Q(\nu, \mu)\circ P(\mu, \nu')} \circ Q(\nu',\mu') \circ E_{n+\hf}(\mu'),
\ea
\label{eq:H-def}
\ee
where 
\be
\ba
P(\mu, \nu)=\frac{1}{2\cosh \frac{\mu-\nu}{2}},\qquad
Q(\nu, \mu)=\frac{1}{2\cosh \frac{\nu-\mu}{2}},\qquad
E_\alpha(\nu) = \re^{\alpha\nu},
\ea
\ee
and the multiplication $\circ$ is defined by
\be
\ba
\cA(\mu, \nu) \circ \cB(\nu, \mu')&:= \int \frac{\rd \nu}{2\pi} \re^{-\frac{\ri k}{4\pi} \nu^2} \cA(\mu, \nu)\cB(\nu, \mu'),\\
\cC(\nu, \mu) \circ \cD(\mu, \nu')&:= \int \frac{\rd \mu}{2\pi} \re^{\frac{\ri k}{4\pi} \mu^2} \cC(\nu, \mu)\cD(\mu, \nu').
\ea
\ee
Note that $\til{H}_{l,a}$ is nothing but the VEV of the Wilson loop in a hook representation 
$R=(a|l)$ in ABJM theory, normalized by the grand partition function
\begin{align}
 \til{H}_{l,a}=\frac{\cW_{(a|l)}^\text{phase}(\kappa,k,0)}{\Xi_0(\kappa,k,0)}.
\end{align}
This quantity was studied in detail in \cite{HHMO}.
On the other hand, $H_{m,n}$ does not have a direct relation to the
Wilson loop in ABJM theory.

\paragraph{Fermionic representation.}
The determinant formula \eqref{eq:W_R-GC-det}
and \eqref{eq:Xiphase-det} have
a natural interpretation in the free fermion language,
as in the case of Wilson loops in ABJM theory \cite{HHMO}. 
Let us consider the free fermion
obeying the anti-commutation relation
\begin{align}
 \{\psi_r,\psi_s^*\}=\cob_{r+s,0}.
\end{align}
One can generalize the free fermion
representation in \cite{HHMO} by
introducing the vacuum with charge $M$
\begin{align}
\begin{aligned}
 |M\ket&=\psi^*_{-\hf}\psi^*_{-1-\hf}\cdots \psi^*_{-M+\hf}|0\ket,\\
\bra M|&=\bra 0|\psi_{\hf}\psi_{1+\hf}\cdots\psi_{M-\hf},
\end{aligned}
\end{align}
where $|0\ket$ is the Fock vacuum annihilated by the positive modes
 \begin{align}
  \psi_r|0\ket=\psi_r^*|0\ket=0,~~~(r>0).
 \end{align}
In other words, $M$ appears as the level of Fermi sea in this representation.
We also introduce the state associated with the modified Frobenius notation
$\h{R}=(\h{a}_1\cdots \h{a}_{r_M}|\h{l}_1\cdots \h{l}_{M+r_M})$
\begin{align}
 \bra \h{R}|=\bra 0|\prod_{i=1}^{r_M} \psi_{\h{a}_i+\hf}^* \prod_{j=1}^{M+r_M}\psi_{\h{l}_j+\hf}.
\end{align}
Then the Wilson loop VEV and the grand partition function
are written as
\begin{align}
 \frac{\cW_R^\text{phase}(\kappa, k, M)}{\Xi_0(\kappa,k,0)}=\bra \h{R}|V|M\ket,\qquad
\frac{\Xi^\text{phase}(\kappa,k,M)}{\Xi_0(\kappa,k,0)}=\bra M|V|M\ket,
\end{align}
where the ``vertex'' $V$ is given by
\begin{align}
V=\exp\left(\sum_{a,l=0}^\infty \til{H}_{l,a}\psi^*_{-l-\hf}\psi_{-a-\hf}\right)
\exp\left(\sum_{a,l=0}^\infty H_{l,a}\psi^*_{-l-\hf}\psi_{a+\hf}\right). 
\end{align}
This is reminiscent of the fermionic representation
of the topological vertex \cite{ADKMV}.

\paragraph{Small $\kappa$ expansions.}
$H_{m,n}$ and $\til{H}_{m,n}$ admit the following small $\kappa$ expansions
\be
\ba
H_{m,n}=\sum_{N=0}^\infty (-\kappa)^N H_{m,n}^{(N)},\qquad
\til{H}_{m,n}=\kappa \sum_{N=0}^\infty (-\kappa)^N \til{H}_{m,n}^{(N)}.
\ea
\ee
The coefficients $\til{H}_{m,n}^{(N)}$
can be written as \cite{HHMO}
\begin{align}
 \til{H}_{m,n}^{(N)}=\re^{\frac{\pi \ri}{k}[n(n+1)-m(m+1)]}\int \frac{dxdy}{(2\pi k)^2}f_{m+\hf}(x)\rho^N(x,y)f_{n+\hf}(y)
\end{align}
where $\rho(x,y)$ denotes the density matrix of ABJM theory
\begin{align}
 \rho(x,y)=\frac{1}{\sqrt{2\cosh\frac{x}{2}}}\frac{1}{2\cosh\frac{x-y}{2k}}
\frac{1}{\sqrt{2\cosh\frac{y}{2}}},
\end{align}
and the function $f_{m+\hf}(x)$ is given by
\begin{align}
 f_{m+\hf}(x)=\frac{\re^{(m+\hf)\frac{x}{k}}}{\sqrt{2\cosh\frac{x}{2}}}.
\label{eq:fm}
\end{align}
As shown in \cite{HHMO}, $\til{H}_{m,n}^{(N)}$ can be computed recursively
by constructing a series of functions $\phi^{(\ell)}(x)$
\begin{align}
 \phi^{(\ell)}(x)=\int \frac{dy}{2\pi k}\rho(x,y)\phi^{(\ell-1)}(y),\qquad
\phi^{(0)}(x)=f_{n+\hf}(x).
\end{align}
Note that the leading term $\til{H}_{m,n}^{(0)}$ in the small
$\kappa$ expansion is given by
\begin{align}
 \til{H}_{m,n}^{(0)}=\frac{\re^{\frac{\pi \ri}{k}[n(n+1)-m(m+1)]}}{2k\cos\frac{\pi(m+n+1)}{k}}.
\end{align}
Similarly, $H_{m,n}^{(N)}$ can be written as \cite{MaMo}
\begin{align}
\begin{aligned}
H_{m,n}^{(0)}&=\frac{\re^{-\frac{\pi\ri}{4}-\frac{\pi\ri}{k}(m-n)^2}}{\rt{k}},\\
 H_{m,n}^{(N)}&=\frac{\re^{-\frac{\pi\ri}{4}-\frac{\pi\ri}{k}[(m+\hf)^2+(n+\hf)^2]}}{\rt{k}}
\int\frac{dxdy}{(2\pi k)^2}g_{m+\hf}(x)\rho^N(x,y)f_{-n-\hf}(y),\quad(N\geq1), 
\end{aligned}
\label{eq:HN-int}
\end{align}
where $f_{-n-\hf}(y)$ is defined in \eqref{eq:fm} and
$g_{m+\hf}(x)$ is given by
\begin{align}
 g_{m+\hf}(x)=\frac{1}{2\cosh\frac{x+2\pi\ri (m+\hf)}{2k}}\frac{1}{\sqrt{2\cosh\frac{x}{2}}}.
\end{align}
Again, $H_{m,n}^{(N)}$ can be computed
recursively by constructing a series of functions. 
Under the exchange of indices $m$ and $n$, 
$H_{m,n}$ is completely symmetric while $\til{H}_{m,n}$ acquires a phase
\begin{align}
H_{n,m}=H_{m,n},\qquad \til{H}_{n,m}=\re^{\frac{2\pi \ri}{k}[m(m+1)-n(n+1)]}\til{H}_{m,n}. 
\end{align} 
Although the symmetry $H_{n,m}=H_{m,n}$ is not manifest in \eqref{eq:HN-int},
we have checked this for various values of $m,n$ and $k$.

From the expression \eqref{eq:W_R-GC-det}, one can see that
the small $\kappa$ expansion of $\cW_R^\text{phase}(\kappa, k, M)$
starts from the term $\kappa^{r_M}$.
For the general representation $R=(a_1\cdots a_r|l_1\cdots l_r)$,
the small $\kappa$ expansion of $\cW_R^\text{phase}(\kappa, k, M)$
takes the following form
\begin{align}
\cW_R^\text{phase}(\kappa, k, M)=
 C_{R}(k,M)\re^{\ri \theta_{R}(k,M)+\ri \theta_0(k,M)}\kappa^{r_M}
\left(1+\sum_{\ell=1}^\infty \Big|W_{R}^{(\ell)}\Big|\re^{\frac{\pi\ri}{2} M\ell}\kappa^{\ell}\right),
\label{eq:WR-expand}
\end{align}
where $C_{R}(k,M)$ is a positive constant
\begin{align}
 C_R(k,M)=\frac{1}{k^{\frac{M}{2}+r_M}}\frac{\prod_{p<p'}2\sin\frac{\pi(\h{l}_{p}-\h{l}_{p'})}{k}
\prod_{q<q'}2\sin\frac{\pi(\h{a}_{q}-\h{a}_{q'})}{k}}{\prod_{p,q}2\cos\frac{\pi(\h{a}_q+\h{l}_p+1)}{k}}. 
\label{eq:CR}
\end{align}
$\theta_0(k,M)$ in \eqref{eq:WR-expand} is the phase 
of the partition function \eqref{eq:th0},
while $\theta_R(k,M)$ in \eqref{eq:WR-expand} is given by
a determinant formula
\begin{align}
 \re^{\ri \theta_{(a_1\cdots a_r|l_1\cdots l_r)}(k,M)}
=\det\Big(\re^{\ri \theta_{(a_i|l_j)}(k,M)}\Big)_{1\leq i,j\leq r},
\label{phase-det}
\end{align}
where the phase of a  hook representation $(a|l)$
is given by
\begin{align}
 \re^{\ri \theta_{(a|l)}(k,M)}
=\exp\left[\frac{\pi \ri}{k}\left(\Big(a+\frac{1-M}{2}\Big)^2
-\Big(l+\frac{1+M}{2}\Big)^2\right)+\frac{\pi\ri M}{2}\right].
\end{align}
This is a generalization of the phase factor of ABJM Wilson
loop found in \cite{HHMO}.
The determinant structure \eqref{phase-det}
is a consequence of the Giambelli formula which we will consider in the next subsection.

\paragraph{Convergence conditions.}
As noticed in \cite{HHMO},
the integral defining $\til{H}_{m,n}$ ($m,n \geq 0$) converges
if $m$ and $n$ satisfy the condition
\begin{align}
 2(m+n+1)<k.
\label{eq:convergence-1}
\end{align}
The convergence condition for $H_{m,n}$ ($m,n \geq 0$) is more subtle.
To see this, we need to go back to the expression \eqref{eq:H-def}. 
From this expression, one obtains the multi-integral representation of $H_{m,n}^{(N)}$ (see \cite{MaMo} for detail),
\be
\ba
H_{m,n}^{(N)}&=\frac{\re^{-\frac{\pi\ri}{4}-\frac{\pi\ri}{k}[(m+\hf)^2+(n+\hf)^2]}}{\rt{k}}
\int \! \prod_{i=1}^N \frac{\rd p_i \rd q_i}{4\pi^2 k}
\re^{-(m+\frac{1}{2})\frac{p_1}{k}} \frac{1}{2\cosh \frac{p_1}{2}}\re^{\frac{\ri}{2\pi k} p_1 q_1}\\
&\times 
\frac{1}{2\cosh \frac{q_1}{2}} \re^{-\frac{\ri}{2\pi k} q_1 p_2} \cdots
\re^{\frac{\ri}{2\pi k} p_N q_N}\frac{1}{2\cosh \frac{q_N}{2}}
\re^{-(n+\frac{1}{2})\frac{q_N}{k}}.
\ea
\ee
It is clear that the integrals over $p_1$ and $q_N$ are convergent only for
\be
m<\frac{k-1}{2},\qquad n<\frac{k-1}{2},
\label{eq:convergence-2}
\ee
due to the exponential factors.
On the other hand, the equation \eqref{eq:HN-int} looks well-defined even for $m \geq (k-1)/2$.
However, the naive application of \eqref{eq:HN-int} does not reproduce the correct
grand partition function \eqref{eq:Xiphase-det} for $M \geq (k+1)/2$,
which must satisfy the Seiberg-like duality:
\be 
\Xi_0(\kappa, k, M)=\Xi_0(\kappa, k, k-M).
\label{eq:seiberg-xi}
\ee
This is already mentioned in \cite{MaMo}. They explain that a reason of this discrepancy
is because the pole at $x=\pi \ri (k-2(m+\frac{1}{2}))$ in \eqref{eq:HN-int}
crosses the real axis for $m>(k-1)/2$. Hence one has to deform the integration contour,
and this leads to an additional contribution after pulling back the contour to the real axis.
We conclude that the expression of $H^{(N)}_{m,n}$ in \eqref{eq:HN-int} is applicable only for
the range \eqref{eq:convergence-2}.
In this work, we will concentrate ourselves to this case.
It is important to extend \eqref{eq:HN-int} for other regimes.

Now we can discuss the convergence condition of
$\cW_R^\text{phase}(\kappa,k,M)$ in \eqref{eq:W_R-GC-det}.
If $r_M=0$, then only the function $H_{m,n}$ appears on the right hand side of \eqref{eq:W_R-GC-det}.
Therefore the convergence condition in this case is
\be
\widehat{l}_p<\frac{k-1}{2},\qquad M-q < \frac{k-1}{2}.
\ee
This gives a restriction on  the allowed size of representations
of the Wilson loops in ABJ theory for a given $k$.
Since $\widehat{l}_p$ is obviously a monotonically decreasing sequence,
the severest condition is
\be
\widehat{l}_1<\frac{k-1}{2},\qquad M-1 < \frac{k-1}{2}.
\ee
Using \eqref{eq:mFrobenius}, this is rewritten as
\be
\lambda_1^T< \frac{k+1}{2}-M, \qquad M<\frac{k+1}{2}.
\label{eq:convergence-3}
\ee
We also have $\lambda_1 \leq M$ because $r_M=0$.

If $r_M>0$ (or $\lambda_1 \geq M+1$), the function $\til{H}_{m,n}$ also
appears in the computation
of $\cW_R^\text{phase}(\kappa,k,M)$ in \eqref{eq:W_R-GC-det}, and we have to impose the additional condition
\be
2(\widehat{a}_q+\widehat{l}_p+1)<k.
\ee
for the convergence of $\til{H}$. The severest condition of this is
\be
2(\widehat{a}_1+\widehat{l}_1+1)<k.
\label{eq:al-cond}
\ee
Since $\widehat{a}_q+\widehat{l}_p=a_q+l_p$,
the convergence condition \eqref{eq:al-cond}
is written as
\be
2(a_1+l_1+1)<k.
\label{eq:convergence-4}
\ee
This condition does not depend on $M$, and thus it is equivalent to
the ABJM case \cite{HHMO}.
Note that $a_1+l_1+1$ in \eqref{eq:convergence-4}
is   the number of boxes in the longest hook of the Young diagram $R$.
In what follows, we will focus on
the representations that satisfy the above conditions.

\subsubsection{Generating function for absolute values}
Since we have determined the phase factors of both the partition function and the Wilson
loop VEVs  explicitly, it is sufficient to consider their absolute
values.
We have already introduced the 
generating function of the absolute value of partition functions
\eqref{eq:Xi0}.
It is also natural to define a generating function for the absolute values of the Wilson loop
VEVs,
\be
\cW_R(\kappa, k, M)=\sum_{N=0}^\infty \kappa^{N} \left| \vev{W_R^{(1/2)}}_{N, N+M, k} \right|.
\label{eq:W_R-abs}
\ee
Also, it is useful to introduce the normalized VEV
in the grand canonical ensemble
\be
\h{\cW}_R(\kappa, k, M)=\frac{\cW_R(\kappa, k, M)}{\Xi_0(\kappa,k,M)}.
\label{eq:hat-WR}
\ee
In the rest of this paper, we will focus on these generating functions.
We will sometimes refer to the generating function $\cW_R(\kappa, k, M)$ of Wilson loop VEVs
simply as Wilson loops, if the meaning is clear from the context.

We find that $\h{\cW}_R(\kappa, k, M)$
satisfies the determinant formula
\be
\h{\cW}_{(a_1\cdots a_r|l_1\cdots l_r)}(\kappa, k, M)
=\det\Big(\h{\cW}_{(a_i|l_j)}(\kappa, k, M)\Big)_{1\leq i,j\leq r}.
\label{eq:Giambelli}
\ee
At the level of the Schur polynomials, such a relation is known as the Giambelli formula.
It is quite surprising that the same formula still holds even after taking the vacuum expectation values!
For the ABJM case $(M=0)$, the formula \eqref{eq:Giambelli} was proved in \cite{HHMO}.
Interestingly, we observe that the formula \eqref{eq:Giambelli} still holds in the ABJ case.
We have checked the formula \eqref{eq:Giambelli}
for various $k,M$ and $R$.

We should mention that
the normalized VEV without taking the absolute values
\begin{align}
 \frac{\cW_R^\text{phase}(\kappa, k, M)}{\Xi^\text{phase}(\kappa,k,M)}
\label{hW-phase}
\end{align}
also satisfies the Giambelli formula.
As a simple check, one can see that
the leading term of \eqref{hW-phase} in the small $\kappa$
expansion indeed satisfies the Giambelli formula\footnote{The Giambelli formula 
for the normalized VEV with phase \eqref{hW-phase} is recently proved in \cite{Mto}.}.

Also, we observe that
the normalized VEV for hook representation
$\h{\cW}_{(a|l)}(\kappa, k, M)$
has a symmetry 
under a generalization of the transpose
of Young diagram
\begin{align}
 \frac{\h{\cW}_{(a|l)}(\kappa, k, M)}{C_{(a|l)}(k,M)}=
 \frac{\h{\cW}_{(l+M|a-M)}(\kappa, k, M)}{C_{(l+M|a-M)}(k,M)},
\label{eq:trpose}
\end{align}
where we have assumed $a\geq M$ (or equivalently $r_M >0$) and the normalization factor $C_{(a|l)}(k,M)$
is given by \eqref{eq:CR}.

The identities
\eqref{eq:Giambelli} and \eqref{eq:trpose}
are the relations among the Wilson
loop VEVs at a fixed $M$.
More interestingly, as we will see below, there are non-trivial relations
connecting the Wilson
loop VEVs at different values of $M$'s. 

From the viewpoint of topological string on local
$\mathbb{P}^1\times\mathbb{P}^1$, the normalized VEV
\eqref{eq:hat-WR}
corresponds to the open string partition function
associated with certain non-compact D-branes \cite{GKM}.
The open topological string partition function can be written as
\begin{align}
Z_\text{open}=\sum_R Z_R \Tr_R V,
\label{eq:Z_open}
\end{align}
where $V$ is an auxiliary $U(\infty)$ matrix, and $R$ runs over all possible
representations of $U(\infty)$.
Then, we have a natural correspondence
between the normalized Wilson loop VEV and the open string amplitude
\begin{align}
 \h{\cW}_R ~~ \longleftrightarrow ~~  Z_R .
 \label{eq:WvsZ}
\end{align}
As we will see in section \ref{sec:open-closed}, 
the above-mentioned relations among the Wilson loop VEVs at
different $M$'s are concrete examples
of the open-closed duality which can be shown very explicitly.

\subsection{The large $N$ limit}
Let us consider the large $N$ limit of Wilson loop VEVs.
It is easy to see that the large $N$ limit corresponds to the large $\mu$ limit in the grand canonical ensemble.
To study the large $\mu$ expansion,
it is useful to consider the ``modified grand potential'' $J_R(\mu,k,M)$ for the generating function \eqref{eq:W_R-abs},
 defined by
\be
\cW_R(\kappa,k,M )=\sum_{n \in \mathbb{Z}} \re^{J_R(\mu+2\pi \ri n, k , M)},\qquad
\kappa=\re^\mu.
\label{eq:mu-sum}
\ee
Obviously, the modified grand potential is different from the standard potential 
$\cJ_R(\mu, k, M)=\log \cW_R(\kappa, k, M)$.
For our purpose, it is more useful to consider the modified grand potential 
than the standard one \cite{HMO2}.
This modified grand potential naturally splits into two parts,
\be
J_R(\mu,k,M )=J(\mu,k,M )+\widehat{J}_R(\mu,k, M),
\label{eq:J_R}
\ee
where $J(\mu,k,M)$ is the modified grand potential of the ABJ grand partition function
in \eqref{eq:Xi0}:
\be
\Xi(\kappa, k , M) = \sum_{n \in \mathbb{Z}} \re^{J(\mu+2\pi \ri n, k , M)}.
\label{eq:Xi-sum}
\ee
The large $\mu$ expansion of $J(\mu,k,M)$ was completely fixed in \cite{HMMO, HoOk}.
In the ABJM case, the large $\mu$ expansion of $\widehat{J}_R(\mu,k,0)$
was also studied in \cite{HHMO} in detail.
The structure of $J_R(\mu,k,M )$ is almost universal.
It is naturally separated into two contributions: a cubic polynomial in $\mu$ and an exponentially suppressed correction.
Thus we can write it as
\be
\ba
J_R(\mu,k,M )=J_R^\text{pert}(\mu,k,M )+J_R^\text{np}(\mu,k,M ),
\ea
\ee
where the first term is the perturbative (polynomial) part and the second term is the exponentially suppressed part.
The perturbative part of the ABJ grand potential $J(\mu,k,M )$ is computed in \cite{MaMo, HoOk}
\be
J^\text{pert}(\mu,k,M )=\frac{C(k)}{3} \mu^3+B(k,M) \mu+A(k,M)
\label{eq:J-pert}
\ee
where
\be
C(k)=\frac{2}{\pi^2 k},\qquad 
B(k,M)=\frac{1}{3k}-\frac{k}{12}+\frac{k}{2}\( \frac{1}{2}-\frac{M}{k} \)^2,
\ee
and
\be
A(k,M)=A_\text{c}(k)-\log Z^\text{CS}(k,M).
\ee
Here $A_\text{c}(k)$ is the so-called the constant map contribution.
Although the expansion of $A_\text{c}(k)$ around $k=0$ or $k=\infty$ has an infinite number of terms,
we can resum this infinite series as a simple integral form \cite{KEK, HO1}
\be
A_\text{c}(k)=-\frac{\zeta(3)}{8\pi^2}k^2+4\int_0^\infty \rd x \, \frac{x}{\re^{2\pi x}-1} \log \( 2\sinh \frac{2\pi x}{k} \).
\ee
In particular, when $k$ is an integer the exact values of $A_\text{c}(k)$ can be written in closed form \cite{HO1}. 

By the same analysis done in \cite{HHMO}, we also find that the perturbative part $J_R^\text{pert}(\mu,k,M )$ for $R$
satisfying the convergence conditions in the previous subsection is generically
written as
\be
J_R^\text{pert}(\mu,k,M )=\frac{C(k)}{3} \mu^3+B_R(k,M) \mu+A_c(k)
+\log A_R(k),
\label{eq:J_R-pert}
\ee
where 
\be
B_R(k,M)=B(k,M)+\frac{2n_R}{k},
\label{eq:B_R}
\ee
with $n_R$ the number of boxes of the corresponding Young diagram for $R$.
When $R$ is a single hook representation $R=(a|l)$, 
the last term of \eqref{eq:J_R-pert} is given by
\be
A_{(a|l)}(k)=\frac{1}{2\sin\frac{2\pi(a+l+1)}{k}\prod_{s=1}^a
2\sin\frac{2\pi s}{k}\prod_{t=1}^l
2\sin\frac{2\pi t}{k}}. 
\label{eq:AR-hook}
\ee
From the Giambelli formula \eqref{eq:Giambelli}, 
the constant $A_R(k)$ for a general representation $R=(a_1\cdots a_r|l_1\cdots l_r)$
is given by
the determinant of \eqref{eq:AR-hook}
\begin{align}
 \begin{aligned}
  A_{(a_1\cdots a_r|l_1\cdots l_r)}(k)=\det\Big(A_{(a_i|l_j)}(k)\Big)_{1\leq i,j\leq r}~.
 \end{aligned}
\end{align}
From \eqref{eq:J_R-pert},
it turns out that the perturbative part of normalized VEV in \eqref{eq:hat-WR}
is independent of $M$
\begin{align}
\h{\cW}_R(\kappa,k,M)^\text{pert}=A_R(k)\re^{\frac{2n_R}{k}\mu}, 
\end{align}
which agrees with the known result of ABJM Wilson loop
\cite{KMSS,HHMO}.

The cubic behavior \eqref{eq:J_R-pert} of modified grand potential
immediately leads to the Airy function behavior in the canonical ensemble \cite{MP}.
Ignoring the non-perturbative corrections in $1/N$, 
the large $N$ behavior of the normalized VEV 
of $1/2$ BPS Wilson loop is given by
\be
\ba
\biggl| \frac{\vev{W_R^{(1/2)}}_{N,N+M,k}}{Z^\text{ABJ}(N,N+M,k)} \biggr| 
\approx A_R(k) \frac{\Ai[ C(k)^{-1/3}(N-B(k,M)-\frac{2n_R}{k})]}{\Ai[ C(k)^{-1/3}(N-B(k,M) )]},
\ea
\ee
where $\approx$ means that all the exponentially suppressed corrections at large $N$ are dropped.

The exponentially suppressed part is generally written as
\be
J_R^\text{np}(\mu,k,M)=\sum_{(\ell,m) \ne (0,0)} f_{\ell, m}(k, M) \re^{-(2\ell+\frac{4m}{k})\mu}.
\ee
where the terms with $\ell=0$ correspond to
 worldsheet instanton corrections,
while the terms with $m=0$ correspond to membrane instanton corrections.
The corrections with $\ell \ne 0$ and $m \ne 0$ are interpreted as bound states of
these two instantons.
It was found in \cite{HMO3} that if one introduces the following ``effective'' chemical potential
\be
\mu_\text{eff}=\mu+\frac{1}{C(k)}\sum_{\ell=1}^\infty a_\ell(k) \re^{-2\ell\mu},
\label{eq:mu-eff}
\ee
then such bound state contributions are absorbed in the worldsheet instanton correction.
The explicit form of the coefficient $a_\ell(k)$ in \eqref{eq:mu-eff}
can be found  in \cite{HMO3,HMMO}.

\section{Exact relations in the ABJ Wilson loops}\label{sec:exact-rel}
In this section, we find the exact relations between the partition function and the 1/2 BPS Wilson loops
in ABJ theory.
As reviewed in the previous section, both can be exactly computed by the localization technique.
Our basic strategy to find such non-trivial relations
is to evaluate these quantities at large $N$ or at finite $N$.

\subsection{The fundamental representation in ABJM theory}

Let us start with the simplest case: the Wilson loop in the
fundamental representation in ABJM theory.
Its large $\mu$ expansion in the grand canonical ensemble was analyzed in \cite{HHMO}.
From \eqref{eq:J_R-pert}, the perturbative part is
given by
\be
\ba
J_{\tableau{1}}^\text{pert}(\mu,k,0)
=\frac{2}{3\pi^2 k}\mu^3+\( \frac{7}{3k}+\frac{k}{24} \)\mu+A_\text{c}(k)-\log \( 2\sin \frac{2\pi}{k} \).
\ea
\ee
Notice that this is related to the perturbative part of the ABJ grand potential with $M=2$ (see \eqref{eq:J-pert}),
\be
J_{\tableau{1}}^\text{pert}(\mu,k,0)=J^\text{pert}(\mu,k,2)+\mu-\log \(2k \cos \frac{\pi}{k} \).
\ee
Surprisingly, we observe that the non-perturbative part is also related to the one for ABJ theory with $M=2$:
\be
J_{\tableau{1}}^\text{np}(\mu,k,0)=J^\text{np}(\mu,k,2).
\ee
We have checked this relation for $k=3,4,6,8,12$ up to the first six terms by using the results in \cite{HHMO}.
Therefore we here conjecture the exact relation
\be
J_{\tableau{1}}(\mu,k,0)=J(\mu,k, 2)+\mu-\log \(2k \cos \frac{\pi}{k} \).
\label{eq:J_(0|0)}
\ee
Then, the generating function
of fundamental Wilson loop  VEVs
becomes
\be
\cW_{\tableau{1}}(\kappa,k,0 )=\sum_{n \in \mathbb{Z}} \re^{J_{\tableau{1}}(\mu+2\pi \ri n,k,0)}=\re^{\mu-\log (2k \cos \frac{\pi}{k} )}
\sum_{n \in \mathbb{Z}} \re^{J(\mu+2\pi \ri n, k ,2)}.
\ee
One can see that the last sum is equal to $\Xi(\kappa,k,2)$
in \eqref{eq:Xi-sum}.
Finally, we arrive at a surprising relation:
\be
\cW_{\tableau{1}}(\kappa,k,0)=\frac{\kappa}{2k \cos \frac{\pi}{k}} \Xi(\kappa, k, 2).
\label{eq:exact-WL-2}
\ee
Namely,
the generating function of Wilson loops in the fundamental
representation in ABJM theory
is equal to
the grand partition of ABJ theory with $M=2$, up to
an overall factor.
Comparing the terms at order $\kappa^N$ on both sides of \eqref{eq:exact-WL-2}, we obtain the relation
for the unnormalized VEV in the canonical picture
\be
\vev{W_{\tableau{1}}^{(1/2)}}_{N,k}=\frac{1}{2k \cos \frac{\pi}{k}} \widehat{Z}(N-1,N+1,k).
\label{eq:exact-WL-3}
\ee
Taking into account the pure CS part and the normalization of the VEV, 
this leads to the exact relation \eqref{eq:exact-WL} mentioned in section \ref{sec:intro}.
Note that for the fundamental representation in ABJM theory the VEV is real-valued~\cite{HHMO}.

In terms of the normalized VEV  in the grand canonical picture \eqref{eq:hat-WR},
one can rewrite \eqref{eq:exact-WL-2} as
\begin{align}
 \h{\cW}_{\tableau{1}}(\kappa,k,0)=
 \frac{\kappa}{2k \cos \frac{\pi}{k}} \frac{\Xi(\kappa, k, 2)}{\Xi(\kappa,k,0)}.
\label{Wfund}
\end{align}
This is an example of the {\it open-closed duality} in \eqref{eq:open-closed} with the identification
\begin{align}
 \h{\cW}_{\tableau{1}}(\kappa,k,0)~\lrya~Z_{\tableau{1}},\qquad
 \Xi(\kappa,k,M)~\lrya~Z_\text{closed}(M),
\end{align}
where $Z_{\tableau{1}}$ is the coefficient corresponding
to $R=\square$ in the expansion of open string
partition function in \eqref{eq:Z_open}.

The relationship \eqref{eq:exact-WL-2} was obtained by the analysis at large $\mu$ (or at large $N$),
but one can check that \eqref{eq:exact-WL-3} is
correct even for finite $N$.
As reviewed in the previous section, the generating function $\cW_R^\text{phase}(\kappa, k, M)$ can be
computed order by order in $\kappa$ by the formula \eqref{eq:W_R-GC-det}. 
The obtained result is easily translated into $\cW_R(\kappa,k,M)$.
In this way, we obtain the following small $\kappa$ expansion at $k=3$, for example,
\be
\ba
\cW_{\tableau{1}}(\kappa,3,0)=\frac{\kappa }{3}+\frac{2 \sqrt{3}-3 }{36}\kappa ^2 +\frac{9-27 \pi +14 \sqrt{3} \pi }{1296 \pi }\kappa ^3+\cO(\kappa^4).
\ea
\label{eq:W3}
\ee
On the other hand, the ABJ grand partition function at $k=3$ was exactly computed in \cite{HoOk},
\be
\Xi(\kappa, 3, 2)=\Xi(\kappa, 3, 1)
=1+\frac{2 \sqrt{3}-3}{12}\kappa+\frac{9-27 \pi +14 \sqrt{3} \pi }{432 \pi }\kappa ^2+\cO(\kappa^3),
\label{eq:Xi32}
\ee
where we used the Seiberg-like duality \eqref{eq:seiberg-xi}.
From \eqref{eq:W3} and \eqref{eq:Xi32}, one can see that
the relation \eqref{eq:exact-WL-2} for $k=3$
\be
\cW_{\tableau{1}}(\kappa,3,0)=\frac{\kappa}{3}\Xi(\kappa, 3, 2),
\ee
is indeed satisfied even for finite $N$.
Similar tests can be done for various $k$.

\subsection{Higher representations in ABJ theory}
Remarkably, the relation \eqref{eq:exact-WL-2}
has a generalization to the 1/2 BPS Wilson loops with higher dimensional representations in ABJ(M) theory.
So far, we do not have
a proof for these relations,
but we have checked them for various $k$ in the small $\kappa$
expansion.

We first find that the representation
$R=[n^n]$
associated with the
$n \times n$ square Young diagram  in ABJM theory
is related to the ABJ partition function with $M=2n$.
Namely, we conjecture
\be
\cW_{[n^n]}(\kappa, k, 0)= \cN_n(k) \kappa^n \Xi(\kappa, k, 2n).
\label{eq:exact-WL-general}
\ee
This is a natural generalization of \eqref{eq:exact-WL-2}, which corresponds to the $n=1$
case of \eqref{eq:exact-WL-general}. 
The  constant $\cN_n(k)$ in \eqref{eq:exact-WL-general}
is given by
\be
\cN_n(k)=\frac{1}{(2k)^n} \frac{\prod_{0\leq a<b\leq n-1}\sin^2\frac{\pi(a-b)}{k}}{\prod_{0\leq a,b\leq n-1}\cos\frac{\pi(a+b+1)}{k}}.
\ee
Again, the normalized VEV takes the form of open-closed duality in
\eqref{eq:open-closed}
\be
\h{\cW}_{[n^n]}(\kappa, k, 0)= \cN_n(k) \kappa^n
\frac{\Xi(\kappa, k, 2n)}{\Xi(\kappa, k, 0)}.
\ee
We have checked the relation \eqref{eq:exact-WL-general} by computing the small $\kappa$ expansion
for various $n$ and $k$ using the formalism in section
\ref{sec:review}.
One can also easily see that the relation
 \eqref{eq:exact-WL-general} is consistent with the perturbative part of the modified grand potential.
Using \eqref{eq:J-pert} and \eqref{eq:J_R-pert}, we find
\be
\ba
J_{[n^n]}^\text{pert}(\mu, k, 0)&=\frac{2}{3\pi^2 k} \mu^3+\( \frac{1}{3k}+\frac{k}{24}+\frac{2n^2}{k} \)\mu+A_{[n^n]}(k,M), \\
J^\text{pert}(\mu, k, 2n)&=\frac{2}{3\pi^2 k} \mu^3+\( \frac{1}{3k}+\frac{k}{24}+\frac{2n^2}{k}-n \)\mu+A(k,2n),
\ea
\ee
which implies
\be
J_{[n^n]}^\text{pert}(\mu, k, 0)=J^\text{pert}(\mu, k, 2n)+n \mu+A_{[n^n]}(k)-A(k,2n).
\ee
This
is nothing but the perturbative pert of the relation \eqref{eq:exact-WL-general}
where the term $\re^{n \mu+A_{[n^n]}(k)-A(k,2n)}$ leads to the factor $\cN_n(k) \kappa^n$.
Of course, to prove the exact relation \eqref{eq:exact-WL-general}, we have to show the equality in the exponentially suppressed
parts:
\be
J_{[n^n]}^\text{np}(\mu, k, 0)=J^\text{np}(\mu, k, 2n).
\label{eq:Jnxn-np}
\ee
Currently, we do not have a direct proof of this relation.
Conversely,  assuming that the relation \eqref{eq:exact-WL-general}
is correct, we can predict $J_{[n^n]}^\text{np}(\mu, k, 0)$ by \eqref{eq:Jnxn-np}
because we already know that the non-perturbative part $J^\text{np}(\mu, k, 2n)$ 
of ABJ grand potential is completely determined by the topological
string free energy on local $\mathbb{P}^1 \times \mathbb{P}^1$ \cite{HoOk}.
In other words, we can predict the large $N$ expansion of the Wilson loop
VEV $|\vev{W_{[n^n]}^{(1/2)}}_{N,k}|$ from the known results
of the ABJ partition function.
Since the square-shape representation is decomposed into the hook representations 
by the Giambelli formula \eqref{eq:Giambelli},
the relation \eqref{eq:exact-WL-general} gives a constraint for these hook representations.

Interestingly, \eqref{eq:exact-WL-general}
can be further generalized to the Wilson loops in ABJ theory.
We find that the 1/2 BPS Wilson loop in ABJ theory with $M=m$
is also related to the ABJ partition function with shifted value of $M=m+2n$;
in this case the Wilson loop is in the representation
$R=[(n+m)^n]$
associated with the 
$n \times (n+m)$ rectangular Young diagram:
\be
\cW_{[(n+m)^n]}(\kappa, k, m) =\cN_{n,m}(k) \kappa^n \Xi(\kappa, k, m+2n).
\label{eq:exact-WL-general-2}
\ee
One can  show that this relation is consistent with the perturbative part of modified grand potential.
In fact, from \eqref{eq:J-pert} and \eqref{eq:J_R-pert}, one finds a simple relation
\be
J_{[(n+m)^n]}^\text{pert}(\mu, k, m)=J^\text{pert}(\mu, k, m+2n)+n \mu+A_{[(n+m)^n]}(k,m)-A(k, m+2n),
\ee
and the  constant $\cN_{n,m}(k)$ in \eqref{eq:exact-WL-general-2}
is thus written as
\be
\cN_{n,m}(k)=\re^{A_{[(n+m)^n]}(k,m)-A(k, m+2n)}.
\ee
We have confirmed the relation \eqref{eq:exact-WL-general-2} by computing the small $\kappa$ expansion
for various $n$, $m$ and $k$.
Note that \eqref{eq:exact-WL-general-2} can also be recast in the form
of open-closed duality \eqref{eq:open-closed}
\begin{align}
 \h{\cW}_{[(n+m)^n]}(\kappa,k,m)=\frac{\cN_{n,m}(k)\kappa^n}{Z^\text{CS}(k,m)}\cdot\frac{\Xi(\kappa,k,m+2n)}{\Xi(\kappa,k,m)}.
\label{eq:open-closed-ABJ}
\end{align}

It is possible to  generalize \eqref{eq:exact-WL-general-2} further.
To sketch this, let us first consider the representation of the form $R=[1^\ell]$ in ABJM theory.
By using \eqref{eq:J_R-pert}, it is easy to see that
\be
J_{[1^\ell]}^\text{pert}(\mu, k, 0)=J_{[1^{\ell-1}]}^\text{pert}(\mu, k ,2)+\mu-\log \( 2\sin \frac{2\pi \ell}{k} \).
\ee
Therefore we expect
\be
\cW_{[1^\ell]}(\kappa, k ,0)=\frac{\kappa}{2\sin \frac{2\pi \ell}{k}} \cW_{[1^{\ell-1}]}(\kappa, k ,2),
\label{eq:exact-WL-vert}
\ee
For the special case $\ell=1$, 
the relation \eqref{eq:exact-WL-vert}  reduces to \eqref{eq:exact-WL-2},
since $[1^0]$ means the trivial representation (no insertion of the loop operator) 
and by definition $\cW_{[1^{0}]}$
is equal to the grand partition function
\be
\cW_{[1^{0}]}(\kappa, k ,M)
=\Xi_0(\kappa, k, M)=Z^\text{CS}(k,M) \Xi(\kappa, k, M).
\ee
We have checked that \eqref{eq:exact-WL-vert} indeed holds by evaluating the first few terms
in the small $\kappa$ expansion. 
The relation \eqref{eq:exact-WL-vert} has an interesting interpretation as an operation on 
the Young diagrams:
The Wilson loop in the representation $[1^{\ell-1}]$ in $M=2$ ABJ theory is
obtained by removing one box from the representation $[1^{\ell}]$ in ABJM theory.
Interestingly, this structure can be generalized for the ABJ Wilson loops, as depicted in Fig.~\ref{fig:Young}.
For example, we find
\be
\ba
\cW_{\tableau{3 3}}(\kappa, k, 1) &\propto \Xi(\kappa, k, 5),\qquad\quad\,\,
\cW_{\tableau{3 3 1}}(\kappa, k, 1) \propto \cW_{\tableau{1}}(\kappa, k, 5), \\
\cW_{\tableau{3 3 2}}(\kappa, k, 1) &\propto \cW_{\tableau{2}}(\kappa, k, 5), \qquad
\cW_{\tableau{3 3 1 1}}(\kappa, k, 1) \propto \cW_{\tableau{1 1}}(\kappa, k, 5).
\ea
\ee
In general,  we find 
$\cW_{R}(\kappa, k , m)\propto \cW_{R'}(\kappa, k , m+2n)$,
where $R'$ is the Young diagram obtained from $R$ 
by removing the $n \times (n+m)$ rectangular part 
from the top of $R$ (see Fig.~\ref{fig:Young}).
In the notation of partitions,
they are related by $R=[(n+m)^{n},R']$.
Thus, the relation in Fig.~\ref{fig:Young} is written as
\be
\cW_{[(n+m)^n,R']}(\kappa, k , m)\propto \cW_{R'}(\kappa, k , m+2n).
\label{eq:WR-WR'}
\ee
We have also checked this relation for several cases
in the small $\kappa$ expansion. 
It would be interesting to find a general proof of \eqref{eq:WR-WR'}.
As we will see in section \ref{sec:open-closed},
this relation \eqref{eq:WR-WR'} 
implies a non-trivial relation \eqref{eq:open-open-2}
for the open string amplitudes  on
local $\mathbb{P}^1\times \mathbb{P}^1$.

\section{More exact results for some special cases}\label{sec:special}
So far, we explored exact relations valid for generic values of $k$.
In this section, we provide some additional results for $k \in 4\mathbb{N}$.
In these special cases, we can further relate the 1/2 BPS Wilson loops
to the grand partition function or its even/odd parity projection.
This allows us to
write the generating function of Wilson loops in closed form. 

As shown in \cite{HMO1}, the grand partition function $\Xi(\kappa,k,M)$ is naturally factorized into 
the ``even'' and ``odd'' parity parts, which we denote by $\Xi_+(\kappa,k,M)$ and $\Xi_-(\kappa,k,M)$,
respectively:
\be
\Xi(\kappa,k,M)=\Xi_+(\kappa, k,M) \Xi_-(\kappa, k , M).
\ee
This factorization was first considered in \cite{HMO1} as a computational tool.
Interestingly, this even/odd parity projection sometimes has a physical meaning, \textit{i.e.}, the functions $\Xi_{\pm}(\kappa,k,M)$
are equivalent to the grand partition function of orientifolded theories \cite{MS1, Okuyama-ori1, Honda-ori, Okuyama-ori2, MS2, MN-ori}. 
Inserting Wilson loops give a new twist in this story.
We find a surprising relation
between the Wilson loop VEV in ABJM theory with $k=4n~(n\in\mathbb{N})$
and $\Xi_\pm(k,M)$ with $(k,M)=(2n,n)$ 
\begin{align}
 \begin{aligned}
  \cW_{R_\text{odd}}(\kappa,4n,0)&=\left(\frac{\ka}{2k}\right)^{[\frac{n}{2}]}
\frac{\prod_{1\leq i<j\leq [\frac{n}{2}]}\sin^2\frac{2\pi(i-j)}{k}}{\prod_{1\leq i,j\leq [\frac{n}{2}] }\cos\frac{\pi(2i+2j-1)}{k}}
\Xi_{+}(\kappa, 2n,n),\\
\cW_{R_\text{even}}(\kappa,4n,0)&=
\left(\frac{\ka}{2k}\right)^{[\frac{n-1}{2}]+1}\frac{\prod_{0\leq i<j\leq [\frac{n-1}{2}]}\sin^2\frac{2\pi(i-j)}{k}}
{\prod_{0\leq i,j\leq [\frac{n-1}{2}]}\cos\frac{\pi(2i+2j+1)}{k}}\Xi_{-}(\kappa, 2n,n).
 \end{aligned}
 \label{eq:W-k4n}
\end{align}
where
\begin{align}
 \begin{aligned}
  R_\text{odd}&=\Big(2\Big[\frac{n}{2}\Big]-1,\cdots,3,1
\Big|2\Big[\frac{n}{2}\Big]-1,\cdots,3,1\Big),\\
R_\text{even}&=\Big(2\Big[\frac{n-1}{2}\Big],\cdots,2,0
\Big|2\Big[\frac{n-1}{2}\Big],\cdots,2,0\Big).
 \end{aligned}
\label{eq:R-oe}
\end{align}
Note that these representations satisfy the convergence conditions in section~\ref{sec:review}.
It is interesting that  restricting the lengths of
arms and legs of a
Young diagram to be even/odd as in \eqref{eq:R-oe}
is related to
the even/odd projection of grand partition functions.

\subsection{$k=4$}
Let us first consider allowed representations at $k=4$ that satisfy \eqref{eq:convergence-3} for $r_M=0$
or \eqref{eq:convergence-3} and \eqref{eq:convergence-4} for $r_M>0$.
It turns out that only the allowed representation is the fundamental representation for $M=0,1$,
and there are no allowed representations for $M \geq 2$.

We want to find out relations between the Wilson loops $\cW_{\tableau{1}}(\kappa, 4,M)$ ($M=0,1$)
and the grand partition function in ABJ theory.
To do so, we look for them by evaluating the small $\kappa$ expansion of the generating function \eqref{eq:W_R-abs}
or the large $\mu$ expansion of the modified grand potential.
There are indeed nice relations!
We find
\be
\ba
\cW_{\tableau{1}}(\kappa, 4,0)&=\frac{\kappa}{4\sqrt{2}} \Xi(\kappa, 4, 2)=\frac{\kappa}{4\sqrt{2}}\Xi_-(\kappa,2,1), \\
\cW_{\tableau{1}}(\kappa, 4,1)&=\frac{1}{2} \Xi_+(\kappa,2,0).
\ea
\ee
The first line are of course the special cases of \eqref{eq:exact-WL-2}, but we further used the exact relation
$\Xi(\kappa, 4, 2)=\Xi_-(\kappa,2,1)$ found in \cite{GHM2}. 
In the present case, the grand partition functions $\Xi_\pm (\kappa,2,0)$ and $\Xi_\pm(\kappa,2,1)$ 
can be written in closed forms \cite{GHM2, Okuyama-ori2}.
Therefore one can compute the above Wilson loops from the known results.
As mentioned above, for $M=2$ there are no representations satisfying the convergence conditions 
\eqref{eq:convergence-3} and \eqref{eq:convergence-4}.
However, if we naively apply the method in section~\ref{sec:review}, we find the relation
\be
\cW_{\tableau{1}}^\text{naive}(\kappa, 4,2)=\frac{1}{2} \Xi(\kappa,4,0)=\frac{1}{2} \Xi_+(\kappa,2,1),
\label{eq:42naive}
\ee
where we have used the relation $\Xi(\kappa,4,0)=\Xi_+(\kappa,2,1)$ \cite{GHM2}.
We should note that the right hand side of \eqref{eq:42naive} might be different from the true function
$\cW_{\tableau{1}}(\kappa, 4,2)$,
since we have computed it using the expression of $H_{m,n}$
in \eqref{eq:HN-int} naively.
It is interesting to explore the true generating function for $M \geq 2$.

\subsection{$k=8$}
At $k=8$, there are several representations satisfying the convergence conditions.
For $M=0,1$, the list of allowed representations is given by
\be
\tableau{1}, \quad \tableau{2},\quad \tableau{1 1},\quad \tableau{3},\quad \tableau{2 1},\quad 
\tableau{1 1 1},\quad \tableau{2 2}.
\ee
For $M=2$, we have
\be
\tableau{1},\quad \tableau{2},\quad \tableau{1 1},\quad \tableau{3},\quad \tableau{2 1},\quad 
\tableau{2 2}.
\ee
For $M=3$, we have
\be
\tableau{1},\quad \tableau{2}, \quad \tableau{3}.
\ee
For $M \geq 4$, there are no allowed representations.

For the fundamental representation, we find the nice exact relations 
\be
\ba
\cW_{\tableau{1}}(\kappa, 8,0)&=\frac{\kappa}{16\cos \frac{\pi}{8}} \Xi(\kappa, 8, 2)=\frac{\kappa}{16\cos \frac{\pi}{8}} \Xi_-(\kappa,4,2), \\
\cW_{\tableau{1}}(\kappa, 8,1)&=\frac{1}{2\sqrt{2}} \Xi_+(\kappa,4,1), \\
\cW_{\tableau{1}}(\kappa, 8,2)&=\frac{1}{4\sqrt{2}} \Xi_-(\kappa,4,0), \\
\cW_{\tableau{1}}(\kappa, 8,3)&=\frac{1}{8\sqrt{2}} \Xi_-(\kappa,4,1), 
\ea
\label{eq:W-k8-fund}
\ee
where we have used $\Xi(\kappa, 8, 2)=\Xi_-(\kappa, 4, 2)$ in the first line \cite{GHM2}.
The functions $\Xi_\pm(\kappa, 4, M)$ ($M=0,1,2$) were exactly computed in \cite{Okuyama-ori2}.  
If we apply the method in section~\ref{sec:review} for $M=4$ naively, we find
\be
\cW_{\tableau{1}}^\text{naive}(\kappa, 8,4)
=\frac{1}{16\sqrt{2} \sin \frac{\pi}{8}} \Xi_+(\kappa,4,2)-\frac{\sin \frac{\pi}{8}}{8} \Xi_-(\kappa,4,2).
\label{eq:W-k8M4-fund}
\ee
This is obtained as follows.
First, in the large $\mu$ limit, we find the closed form of the grand potential
\begin{align}
J_{\tableau{1}}^\text{naive}(\mu,8,4)=J_{-}(\mu, 4,2)+\frac{\mu}{2}+\log\( \frac{1}{16} \sin \frac{\pi}{8} \)
-\text{arcsinh}(2\re^{-\frac{\mu}{2}}),
\label{eq:J-k8M4-fund}
\end{align}
where $J_-(\mu,4,2)$ was computed in \cite{Okuyama-ori2}.
Once the exact form of the modified grand potential is found,
we can write down the generating function
by summing over the $2\pi\ri$-shift of $\mu$, as in \eqref{eq:mu-sum}.
The result is written in terms of a sum of theta functions.
In the present case, the non-trivial part in \eqref{eq:J-k8M4-fund} is encoded in $J_-(\mu,4,2)$, and thus 
it is expected that the generating function
is related to $\Xi_-(\kappa,4,2)$.
Indeed, it is straightforward to show the following expression
\be
\ba
\cW_{\tableau{1}}^\text{naive}(\kappa,8,4)
=\frac{\sin \frac{\pi}{8}}{16} \( -2\Xi_-(\kappa,4,2)+\sqrt{4+\kappa} \sum_{n \in \mathbb{Z}} (-1)^n \re^{J_-(\mu+2\pi \ri n,4,2)} \), 
\ea
\label{eq:J-k8M4}
\ee
The second term in parenthesis is further written in terms of $\Xi_+(\kappa, 4,2)$.
To see this, we use the result in \cite{Okuyama-ori2}.
As found in \cite{Okuyama-ori2}, the difference of the grand potential is
\be
J_+(\mu,4,2)-J_-(\mu,4,2)=\frac{\mu}{2}+\frac{1}{2}\log 2+2\log \( \sin \frac{\pi}{8} \) +\frac{1}{2} \log (1+4\re^{-\mu}).
\ee
Hence we get
\be
\Xi_+(\kappa,4,2)=\sqrt{2} \sin^2 \frac{\pi}{8} \, \sqrt{4+\kappa} \sum_{n \in \mathbb{Z}} (-1)^n \re^{J_-(\mu+2\pi \ri n,4,2)}.
\ee
We finally arrive at the last line in \eqref{eq:W-k8M4-fund}.
However, one should keep in mind that the naive function $\cW_{\tableau{1}}^\text{naive}(\kappa,8,4)$ may
not be equal to the correct function $\cW_{\tableau{1}}(\kappa,8,4)$.

For $M=0$, we further obtain the relationships for higher representations
\be
\ba
\cW_{\tableau{2}}(\kappa,8,0 )&=\cW_{\tableau{1 1}}(\kappa,8,0 )=\frac{\kappa}{8\sqrt{2}} \Xi_-(\kappa,4,0), \\
\cW_{\tableau{2 1}}(\kappa,8,0)&=\frac{\kappa}{16\sin \frac{\pi}{8} } \Xi_+( \kappa ,4,2), \\
\cW_{\tableau{3}}(\kappa, 8,0)&=\cW_{\tableau{1 1 1}}(\kappa,8,0 )
=\kappa \left[ \frac{1}{16\sqrt{2} \sin \frac{\pi}{8}} \Xi_+(\kappa,4,2)+\frac{\sin \frac{\pi}{8}}{8} \Xi_-(\kappa,4,2) \right].
\ea
\ee
The last equation is obtained from the exact form of the modified grand potential
\be
J_{\tableau{3}}(\mu,8,0)=J_{-}(\mu, 4,2)+\frac{3\mu}{2}+\log\( \frac{1}{16} \sin \frac{\pi}{8} \)
+\text{arcsinh}(2\re^{-\frac{\mu}{2}}).
\label{eq:J-k8M0-3}
\ee
This is almost identical to \eqref{eq:J-k8M4-fund}, and one can repeat the same computation
as above.

Now let us recall that the normalized VEV for $R=\tableau{2 2}=(10|10)$ is given by the Giambelli formula
\be
\h{\cW}_{\tableau{2 2}}=\det \begin{pmatrix}
\h{\cW}_{\tableau{2 1}} & \h{\cW}_{\tableau{2}} \\
\h{\cW}_{\tableau{1 1}} & \h{\cW}_{\tableau{1}}
\end{pmatrix}.
\ee
This is rewritten as
a relation for the unnormalized Wilson loops
\be
\Xi(\kappa,k,M)\cW_{\tableau{2 2}}(\kappa,k,M)=\cW_{\tableau{2 1}}(\kappa,k,M)\cW_{\tableau{1}}(\kappa,k,M)
-\cW_{\tableau{2}}(\kappa,k,M)\cW_{\tableau{1 1}}(\kappa,k,M).
\ee
Furthermore, using the relation \eqref{eq:exact-WL-general}, we have
\be
\cW_{\tableau{2 2}}(\kappa,k,0)
=\frac{\sin^2 \frac{\pi}{k}}{4k^2 \cos^2 \frac{2\pi}{k} \cos \frac{\pi}{k} \cos \frac{3\pi}{k}} \kappa^2 \Xi(\kappa,k,4).
\ee
Combining all these relations, we finally arrive at the following functional equation among
the ABJ grand partition functions:
\be
(\sqrt{2}-1)\Xi(\kappa,8,0)\Xi(\kappa,8,4)=\sqrt{2}\, \Xi(\kappa,4,2)-\Xi_-(\kappa,4,0)^2.
\ee
This is a highly non-trivial consequence of the above consideration.
This can be confirmed by using the results in \cite{HoOk,Okuyama-ori2}.
We have indeed checked it by evaluating the small $\kappa$ expansion of the grand partition function
up to $\kappa^{20}$.

It is not easy to find out a pattern when the generating functions
of Wilson loops are related to the grand partition functions,
but we get some relations for $M \geq 1$.
For $M=1$,
\be
\ba
\cW_{\tableau{2}}(\kappa, 8,1)&=\frac{\sqrt{2}-1}{32} \kappa \Xi(\kappa, 8, 3) ,\\
\cW_{\tableau{3}}(\kappa, 8,1)&=\cW_{\tableau{2 1}}(\kappa, 8,1)=\frac{\kappa}{16\sqrt{2}} \Xi_-(\kappa, 4, 1), \\
\cW_{\tableau{1 1 1}}(\kappa,8,1)&=\frac{1}{2\sqrt{2}} \Xi_+(\kappa,4,1)+\frac{\kappa}{32} \Xi_-(\kappa, 4, 1), \\
\ea
\ee
For $M=2$,
\be
\ba
\cW_{\tableau{2}}(\kappa, 8, 2)&=\sqrt{2}\cW_{\tableau{1}}^\text{naive}(\kappa,8,4), \qquad\qquad
\cW_{\tableau{1 1}}(\kappa, 8,2)=\sqrt{2}\cW_{\tableau{3}}(\kappa, 8, 0), \\
\cW_{\tableau{3}}(\kappa, 8, 2)&= \frac{\sqrt{2}-1}{64} \kappa \Xi(\kappa, 8, 4), \qquad
\cW_{\tableau{2 1}}(\kappa, 8, 2)=\frac{1}{4\sqrt{2}} \Xi_+(\kappa, 4, 0), 
\ea
\ee
For $M=3$,
\be
\ba
\cW_{\tableau{3}}(\kappa,8,3)=\frac{1}{8} \Xi_+(\kappa,4,1)-\frac{1}{8\sqrt{2}} \Xi_-(\kappa, 4, 1) , 
\ea
\ee
The expressions of $\cW_{\tableau{1 1 1}}(\kappa,8,1)$ and $\cW_{\tableau{3}}(\kappa,8,3)$ are
obtained by using the following results
of the closed form of modified grand potentials
\begin{align}
 \begin{aligned}
  J_{\tableau{1 1 1}}(\mu,8,1)&=J_{+}(\mu, 4,1)+\frac{\mu}{2}-3\log 2
+\hf\log(1-8\re^{-\mu})+\text{arctanh}(2\rt{2}\re^{-\frac{\mu}{2}}),\\
 J_{\tableau{3}}(\mu,8,3)&=J_{-}(\mu, 4,1)+\frac{\mu}{2}-5\log 2
+\hf\log(1-8\re^{-\mu})-\text{arctanh}(2\rt{2}\re^{-\frac{\mu}{2}}).
 \end{aligned}
\end{align}
It would be interesting to clarify when we can relate Wilson
loops to
the grand partition functions more systematically.

\subsection{$k=12$}
Since many representations are allowed at $k=12$, we do not write them down here.
Again, we find non-trivial relations between Wilson loops and grand partition functions.
In the ABJM case, from  \eqref{eq:exact-WL-2} and \eqref{eq:exact-WL-general}
we have
\be
\ba
\cW_{\tableau{1}}(\kappa,12, 0)&=\frac{\kappa}{6(\sqrt{2}+\sqrt{6})} \Xi(\kappa,12,2), \\
\cW_{\tableau{2 2}}(\kappa,12, 0)&=\frac{\kappa^2 }{432(3\sqrt{3}+5)} \Xi(\kappa,12,4).
\ea
\ee
and from  \eqref{eq:W-k4n} we find
\be
\ba
\cW_{\tableau{2 1}}(\kappa,12, 0)&=\frac{\kappa}{12\sqrt{2}} \Xi_+(\kappa,6,3), \\
\cW_{\tableau{3 2 1}}(\kappa,12, 0)&=\frac{\kappa^2}{288}\Xi_-(\kappa,6,3).
\ea
\ee
By using the Giambelli identity, this leads to the determinant identity
\be
\det \begin{pmatrix}
\cW_{\tableau{3 1 1}}(\kappa,12,0) & \cW_{\tableau{3}}(\kappa,12,0) \\
\cW_{\tableau{1 1 1}}(\kappa,12,0) & \cW_{\tableau{1}}(\kappa,12,0)
\end{pmatrix}
=\frac{\kappa^2}{288} \Xi(\kappa,12,0)\Xi_-(\kappa,6,3).
\ee
We also find
\be
\cW_{\tableau{3}}(\kappa, 12,1)=\frac{1}{\sqrt{3}}\cW_{\tableau{2 1}}(\kappa,12,1)
=\frac{\sqrt{3}-1}{144} \kappa \Xi_-(\kappa,6,2).
\ee

\section{Open-closed duality for topological string amplitudes}\label{sec:open-closed}

\subsection{The fundamental representation}
The fact that the ABJ Wilson loops are related to the ABJ partition function
implies that the open topological invariants and closed topological invariants
are interrelated.
Here we explicitly show that this is indeed the case  for the fundamental representation.

Let us consider the relation \eqref{eq:J_(0|0)}
between $J_{\tableau{1}}(\mu,k,0)$ and
 $J(\mu,k,2)$.
As found in \cite{HoOk}, the worldsheet instanton part in the ABJ grand potential is
given by the standard (un-refined) topological string free energy
on local $\mathbb{P}^1\times \mathbb{P}^1$
\be
\ba
F_\text{top}(Q_1,Q_2;g_s)&=\log Z_\text{closed}(Q_1,Q_2;g_s)\\
&=\sum_{g \geq 0} \sum_{w \geq 0} \sum_{d_1,d_2 \geq 0} \frac{1}{w} n_{g}^{d_1,d_2} \( 2\sin \frac{w g_s}{2} \)^{2g-2} Q_1^{wd_1}Q_2^{wd_2},
\ea
\label{eq:WS}
\ee
where $n_g^{d_1,d_2}$ is an integer, called the Gopakumar-Vafa (GV) invariant, and the string coupling $g_s$ is related to the Chern-Simons level $k$ by 
\be
g_s=\frac{4\pi}{k}.
\ee
The K\"ahler moduli are also related to the ``effective'' chemical potential $\mu_\text{eff}$
in \eqref{eq:mu-eff} by
\be
T_1=\frac{4\mu_\text{eff}}{k}+2\pi \ri \( \frac{1}{2}-\frac{M}{k} \),\qquad
T_2=\frac{4\mu_\text{eff}}{k}-2\pi \ri \( \frac{1}{2}-\frac{M}{k} \),
\label{eq:moduli}
\ee
and
\be
Q_1=\re^{-T_1}=Qq^M, \qquad Q_2=\re^{-T_2}=Qq^{-M}.
\label{eq:Q1Q2-M}
\ee
Here we have also defined
\begin{align}
 Q=-\re^{-\frac{4\mu_\text{eff}}{k}},\qquad
q=\re^{\frac{2\pi\ri}{k}}=\re^{\frac{\ri g_s}{2}}.
\end{align}
Note that the relation 
between $\mu_\text{eff}$ 
and $\mu$ in \eqref{eq:mu-eff}
is interpreted in \cite{HMMO} as the quantum mirror map
in the topological string \cite{ACDKV}. 

On the other hand, $\widehat{J}_{\tableau{1}}(\mu,k,0)$
is related to the open string amplitude.
According to \cite{HHMO}, the complete large $\mu$ expansion of $\widehat{J}_{\tableau{1}}(\mu,k,0)$ is given by
\be
\widehat{J}_{\tableau{1}}(\mu,k,0)=\frac{2\mu_\text{eff}}{k}-\log\( 2\sin \frac{2\pi}{k} \)
+\log Z_{\tableau{1}}(Q,Q;g_s),
\ee
where $Z_{\tableau{1}}(Q,Q;g_s)$
is the open string amplitude in the fundamental representation
for the ``diagonal'' local  $\mathbb{P}^1\times \mathbb{P}^1$ ({\it i.e.} $Q_1=Q_2=Q$)
\be
Z_{\tableau{1}}(Q,Q;g_s)=\sum_{g \geq 0}\sum_{d \geq 0} (-1)^g n_{g,(1)}^d \(2 \sin \frac{g_s}{2} \)^{2g} Q^d,
\ee
where $n_{g,(1)}^d$ is the open GV invariant corresponding to the fundamental representation.
We define $n^0_{0,(1)}=1$.

Now, plugging these results into \eqref{eq:J_(0|0)} and comparing the worldsheet instanton 
part on both sides of \eqref{eq:J_(0|0)}, we get
\be
\ba
F_\text{top}(Q,Q;g_s)
+\log Z_{\tableau{1}}(Q,Q;g_s)=F_\text{top}(Qq^2,Qq^{-2};g_s).
\ea
\label{eq:F-open-closed}
\ee
This can also be written as
\be
Z_{\tableau{1}}(Q,Q;g_s)
=\frac{Z_\text{closed}( Qq^{2},  Qq^{-2}; g_s)}{Z_\text{closed}(Q, Q; g_s)}.
\label{eq:open-closed-2}
\ee 
This is the main result in this subsection.
Using the explicit values of the closed GV invariants $n_g^{(d_1,d_2)}$ in \cite{AMV} 
and open GV invariants $n_{g,(1)}^d$ in \cite{GKM},
one can confirm that the relation \eqref{eq:open-closed-2} indeed holds.
The relation \eqref{eq:open-closed-2} uniquely fixes the open GV invariants from the closed ones.
For instance,  we find the following non-trivial
relation between open GV invariants and closed GV invariants
\be
\ba
n_{0,(1)}^1&=-n_0^{0,1},\\
n_{0,(1)}^2&=\frac{1}{2}n_0^{0,1}(n_0^{0,1}-1), \\
n_{0,(1)}^3&=-\frac{1}{6}n_0^{0,1}(n_0^{0,1}-1)(n_0^{0,1}-2)-n_0^{1,2}, \\
n_{0,(1)}^4&=\frac{1}{24}n_0^{0,1}(n_0^{0,1}-1)(n_0^{0,1}-2)(n_0^{0,1}-3)+n_0^{0,1}n_0^{1,2}-4n_0^{1,3},\\
n_{1,(1)}^4&=-n_0^{(1,3)}.
\ea
\ee
As explained in \cite{ADKMV}, the open topological string amplitude is generically related to
the closed topological string amplitude.
The effect of the open string shifts the moduli of the closed string
\eqref{eq:open-closed}.
Our exact relation \eqref{eq:open-closed-2}
indeed reflects this fact, 
and it is a concrete example of the open-closed duality
\eqref{eq:open-closed}.

Before closing this subsection, let us comment on the membrane instanton corrections.
As shown in \cite{HMMO}, the membrane instanton corrections to the grand potential
is given by the topological string free energy in the so-called Nekrasov-Shatashvili limit.
Let $F_\text{NS}(T_1, T_2; \hbar)$ be the Nekrasov-Shatashvili free energy on local 
$\mathbb{P}^1 \times \mathbb{P}^1$. 
In our context, we identify $\hbar=\pi k=4\pi^2/g_s$ (see \cite{HMMO}).
Then the membrane instanton correction to the ABJ 
grand potential is written as
\be
\frac{\pd}{\pd g_s} \left[ g_s F_\text{NS}\(\frac{2\pi T_1}{g_s},\frac{2\pi T_2}{g_s};\frac{4\pi^2}{g_s}\) \right].
\ee
Note that the coupling dependence in the argument is $1/g_s$, not $g_s$.
This reflects the fact that the membrane instanton corrections are indeed
non-perturbative corrections in $g_s$.
They are not visible in the 't Hooft limit $\mu \to \infty$ and $g_s \to 0$ with $g_s \mu$ kept finite.
In this limit, only the worldsheet instanton correction \eqref{eq:WS} survives.
We notice that the K\"ahler moduli \eqref{eq:moduli} satisfy the relation
\be
\frac{2\pi T_{1,2}^{(M)}}{g_s}-\frac{2\pi T_{1,2}^{(M=0)}}{g_s}=\pm \pi \ri M.
\ee
One can see that the membrane instanton factor $\re^{-2\pi T_{1,2}/g_s}$
is  identical for $M=0$ and $M=2$ (or more generally for even $M$).
This implies that the membrane instanton corrections on the right hand side of \eqref{Wfund}
precisely cancel between the numerator and the denominator.
As a result, the normalized VEV
in the grand canonical picture
$\widehat{\cW}_{\tableau{1}}(\kappa, k, 0)$ does not receive ``pure'' membrane instanton corrections, 
except for the bound state corrections coming from the replacement  $\mu\to\mu_\text{eff}$.
This explains
the absence of ``pure'' membrane instanton corrections observed in \cite{HHMO}.

\subsection{Higher representations}
The relation \eqref{eq:open-closed-2}
for 
the fundamental representation  can be generalized to the representation
$R=[(n+m)^n]$,
by translating the relation \eqref{eq:open-closed-ABJ}
in ABJ theory
into the language of the topological string
on local $\mathbb{P}^1\times \mathbb{P}^1$    
 \begin{align}
  Z_{[(n+m)^n]}(Qq^{m},Qq^{-m};g_s)=\frac{Z_{\text{closed}}(Qq^{m+2n},Qq^{-m-2n};g_s)}{Z_{\text{closed}}(Qq^{m},Qq^{-m};g_s)}.
\label{eq:open-closed-M}
 \end{align}
In this relation,
the closed string side is given by \eqref{eq:WS},
while the open string amplitudes $Z_R(Q_1,Q_2;g_s)$
for the general representation $R$ can be
computed by the technique of topological vertex \cite{Aganagic:2003db}.\footnote{%
The open string amplitudes in this subsection are obtained
by using a {\tt Mathematica} program written by Marcos Mari\~no. We would like to thank him for 
sharing the program with us.}
Thus, in this case
we can explicitly check the open-closed duality \eqref{eq:open-closed-M}
of topological string
amplitudes predicted by the analysis of ABJ Wilson loops.
Note that, when comparing the open and closed string amplitudes
in  \eqref{eq:open-closed-M}, we should normalize the
open string amplitude as
\begin{align}
 Z_R(0,0;g_s)=1.
 \label{eq:normal-open}
\end{align}
Indeed, we find a complete agreement 
between both sides of \eqref{eq:open-closed-M} for various cases. 
For $m=0$, we find
\begin{align}
  \begin{aligned}
     &Z_{\tableau{1}}(Q,Q;g_s)=\frac{Z_\text{closed}(Qq^{2},Qq^{-2};g_s)}{Z_\text{closed}(Q,Q;g_s)}\\
   &=1+2 Q + 3 Q^2 + 10 Q^3 + \Big( 8 (q^2+q^{-2})+33\Big) Q^4 +
   \mathcal{O}(Q^5),\\
  &Z_{\tableau{2 2}}(Q,Q;g_s)=\frac{Z_\text{closed}(Qq^{4},Qq^{-4};g_s)}{Z_\text{closed}(Q,Q;g_s)}\\
&=
1+\Big(2(q^2+q^{-2})+4\Big)Q+\Big(3 (q^4+q^{-4})+8 (q^2+q^{-2})+14\Big)Q^2+\mathcal{O}(Q^3),\\
&Z_{\tableau{3 3 3}}(Q,Q;g_s)=\frac{Z_\text{closed}(Qq^{6},Qq^{-6};g_s)}{Z_\text{closed}(Q,Q;g_s)}\\
&=
1+\Big(2 (q^4+q^{-4})+4 (q^2+q^{-2})+6\Big)Q\\
&\quad+\Big(3 (q^8+q^{-8})+8 (q^6+q^{-6})+22 (q^4+q^{-4})+32 (q^2+q^{-2})+41\Big)Q^2+\mathcal{O}(Q^3).
  \end{aligned}
\end{align}
For $m=1$, we find
\begin{align}
 \begin{aligned}
  &Z_{\tableau{2}}(Qq,Qq^{-1};g_s)=\frac{Z_\text{closed}(Qq^{3},Qq^{-3};g_s)}
{Z_\text{closed}(Qq,Qq^{-1};g_s)}\\
&=
  1+2(q+q^{-1})Q+\Big(3 (q^2+q^{-2})+4\Big)Q^2+
\Big(4 (q^3+q^{-3})+12 (q+q^{-1})\Big)Q^3+
  \mathcal{O}(Q^4),\\
&Z_{\tableau{3 3}}(Qq,Qq^{-1};g_s)=\frac{Z_\text{closed}(Qq^{5},Qq^{-5};g_s)}
{Z_\text{closed}(Qq,Qq^{-1};g_s)}\\
&=
1+\Big(2 (q^3+q^{-3})+4( q+q^{-1})\Big)Q\\
&\quad+\Big(3 (q^6+q^{-6})+8 (q^4+q^{-4})+18 (q^2+q^{-2})+20\Big)Q^2+\mathcal{O}(Q^3).
 \end{aligned}
\end{align}
For $m=2$, we find
\begin{align}
 \begin{aligned}
  &Z_{\tableau{3}}(Qq^{2},Qq^{-2};g_s)=\frac{Z_\text{closed}(Qq^{4},Qq^{-4};g_s)}
{Z_\text{closed}(Qq^{2},Qq^{-2};g_s)}\\
&=
1+2(q^2+q^{-2}+1)Q+\Big(3 (q^4+q^{-4})+4( q^2+q^{-2})+7\Big)Q^2+\mathcal{O}(Q^3),\\
&Z_{\tableau{4 4}}(Qq^{2},Qq^{-2};g_s)=\frac{Z_\text{closed}(Qq^{6},Qq^{-6};g_s)}
{Z_\text{closed}(Qq^{2},Qq^{-2};g_s)}\\
&=
1+\Big(2 (q^4+q^{-4})+4 (q^2+q^{-2})+4)\Big)Q\\
&\quad+\Big(3 (q^8+q^{-8})+8( q^6+q^{-6})+18( q^4+q^{-4})+24( q^2+q^{-2})+30\Big)Q^2+\mathcal{O}(Q^3).
 \end{aligned}
\end{align}
One can continue this check for other $m$'s and representations.
We should stress that the open string side and the closed
string side can be computed independently, and we find 
a perfect agreement on both sides in a quite non-trivial way.

Moreover, the claim \eqref{eq:WR-WR'} 
in ABJ theory (see also Fig.~\ref{fig:Young}) is translated as
\be
\ba
 &Z_{[(n+m)^n,R]}(Qq^{m},Qq^{-m};g_s)Z_{\text{closed}}(Qq^{m},Qq^{-m};g_s) \\
 &\quad=Z_{R}(Qq^{m+2n},Qq^{-m-2n};g_s)Z_{\text{closed}}(Qq^{m+2n},Qq^{-m-2n};g_s),
\label{eq:open-open}
\ea
\ee
where we denoted $R'$ in \eqref{eq:WR-WR'}  by $R$ for notational simplicity.
Using \eqref{eq:open-closed-M},
we can further rewrite \eqref{eq:open-open}
 as an ``open-open'' relation
\be
\frac{Z_{[(n+m)^n,R]}(Qq^{m},Qq^{-m};g_s)}{Z_{[(n+m)^n]}(Qq^{m},Qq^{-m};g_s)}
=Z_{R}(Qq^{m+2n},Qq^{-m-2n};g_s).
\label{eq:open-open-2}
\ee
We have checked this relation for several examples
for the first few terms in the small $Q$ expansion.
It would be interesting to
find a proof of this conjecture \eqref{eq:open-open-2}
from the topological vertex.

Finally, the transpose formula \eqref{eq:trpose} predicts the relation
\be
Z_{(a|l)}(Qq^{M}, Qq^{-M};g_s)=Z_{(l+M|a-M)}(Qq^{M},Qq^{-M};g_s).
\ee
We have indeed confirmed this for various $a$, $l$ and $M$.

\subsection{Comment on Seiberg-like duality}
We would like to understand the mapping of the 1/2 BPS Wilson loops in ABJ theory
under the Seiberg-like duality $(k,M)\to (k,k-M)$.
However, our method in section \ref{sec:review}
is applicable
only in a limited range of $M$.
Therefore, currently
we do not have a computational method to directly study the
Seiberg-like duality of Wilson loops in ABJ theory.
On the other hand, the topological string description of the ABJ Wilson loops does not
seem to have any restriction.
Thus, this relation
helps us to predict the behavior of the ABJ Wilson loops under the Seiberg-like duality.

From the viewpoint of 
the topological string on local $\mathbb{P}^1\times \mathbb{P}^1$,
the Seiberg-like duality of ABJ theory corresponds 
to the exchange of two $\mathbb{P}^1$'s of local $\mathbb{P}^1\times \mathbb{P}^1$ \cite{HoOk}.
Under this exchange of two $\mathbb{P}^1$'s,
the representation $R$ of the open string amplitude
 simply becomes
its transpose $R^T$
\begin{align}
 Z_R(Q_1,Q_2;g_s)=Z_{R^T}(Q_2,Q_1;g_s),
 \label{eq:ZR-trans}
\end{align}
where we have to use the normalization \eqref{eq:normal-open} on both sides.
 We do not have a general proof of this
relation, but we have checked this behavior for various
representations.
From the correspondence between $\h{\cW}_R$
and $Z_R$ in \eqref{eq:WvsZ},
the property \eqref{eq:ZR-trans}
of open string amplitudes
implies that the
 transformation of the Wilson loops in ABJ theory
 under
 the Seiberg-like duality is simply given by
 the transpose of Young diagram, up to an overall factor
 \begin{align}
  \h{\cW}_R(\kappa,k,M)\propto\h{\cW}_{R^T}(\kappa,k,k-M).
 \end{align}
 Using the symmetry of grand partition function
 \eqref{eq:seiberg-xi},
 the Seiberg-like duality
 of the unnormalized  Wilson loops is again
 given by
 the transpose of $R$
\begin{align}
 \cW_R(\kappa,k,M)\propto\cW_{R^T}(\kappa,k,k-M).
 \label{eq:seiberg-W}
\end{align}
It would be interesting to confirm \eqref{eq:seiberg-W} directly from the ABJ matrix model.
\section{Conclusion and future directions}\label{sec:conclusion}
In this work, we demonstrated many exact results for the 1/2 BPS Wilson loops in ABJ theory.
The most general one is \eqref{eq:WR-WR'}, shown in Fig. \ref{fig:Young}.
In particular, we found novel and surprising relations between the partition function 
and the VEVs of the 1/2 BPS Wilson loops.
These relations are naturally interpreted as the open-closed duality.
Indeed, we found a non-trivial relation between the open topological string partition function
and the closed one.

There are many issues to be understood more deeply.
First of all, it is desirable to prove the conjecture \eqref{eq:exact-WL-general-2}
and its generalization in \eqref{eq:WR-WR'}.
There are several possible routes to do that.
One approach is to rewrite the original matrix integral directly.
For instance, 3d mirror symmetry in Chern-Simons-matter theories was successfully shown in this approach \cite{KWY2}.
However, it seems to be technically difficult to do it in our case because of the insertion of the
supersymmetric Schur polynomial.
The second one is to show the equality of the modified grand potential.
Since we know that the modified grand potential at large $\mu$ is related to the topological string free energy,
the problem is equivalent to show the relations in the open and closed topological strings,
as we have seen in the previous section.
The third one is to use the free Fermion representation in subsection~\ref{sec:Fermi-gas}.
In this picture, the ABJ Wilson loops is understood as excitations over the ``dressed'' vacuum $|M\ket$. 
Fig.~\ref{fig:Young} is then interpreted as excitations over the two different vacua $|m \ket$ and $|m+2n \ket$.
In this approach, the problem is mapped to show the equality in a purely algebraic way of the Fermionic operators.
It would be interesting to consider a proof of \eqref{eq:exact-WL-general-2} along these lines.

From the behavior of open topological string amplitudes
 in local $\mathbb{P}^1\times \mathbb{P}^1$,
 we conjecture that
 the Seiberg-like duality of 1/2 BPS Wilson loops
in ABJ theory is simply given by the transpose of Young diagram
 \eqref{eq:seiberg-W}.
In \cite{HNS}, the Seiberg-like duality 
of the winding Wilson loops,
which is a linear combination of hook representations, in ABJ theory was studied. It was found that the winding Wilson
loop is mapped to itself  (up to a sign)
 under the Seiberg-like duality,
 which is consistent with our conjecture
\eqref{eq:seiberg-W}.
As far as we know, there seem to be no results on the Seiberg-like duality for the general representations in ABJ theory. 
In \cite{Kapustin:2013hpk},
it is found that the Wilson loops in $\mathcal{N}=2$
Chern-Simons-matter theories
transform under the Seiberg-like duality in a very intricate way.
We expect that the Seiberg-like
duality for the Wilson loops in $\mathcal{N}=6$ 
ABJ theory has a simpler structure than
the $\mathcal{N}=2$ cases studied in \cite{Kapustin:2013hpk},
and indeed our conjecture is basically the same as the mapping of
Wilson loops in pure CS theory under the level-rank duality. 
To study the Seiberg-like duality of ABJ Wilson loops further,
we need to resolve the issue of the computation of $H_{m,n}$.
As seen in section~\ref{sec:review}, the formalism used here is applicable only for the range \eqref{eq:convergence-2},
though the original definition \eqref{eq:H-def} seems to be
well-defined for any $m$ and $n$.
It is desirable to extend the formalism in \cite{MaMo} to arbitrary $m$ and  $n$.
If this can be done, we can compute the Wilson loop VEVs for the whole range
$0\leq M \leq k$,
and study the Seiberg-like duality.
It would be very important to develop
a technique to compute $H_{m,n}$ beyond the bound
\eqref{eq:convergence-2}.

The fact that  the 1/2 BPS Wilson loops is related to the ABJ grand partition
function suggests that the generating function $\cW_R(\kappa, k, M)$ may be also regarded as 
a grand partition
function of an non-interacting Fermi-gas.
For example, the equation \eqref{eq:exact-WL-2} can be rewritten as
\be
\cW_{\tableau{1}}(\kappa, k, 0)=\frac{\kappa}{2k \cos \frac{\pi}{k}} \prod_{n=0}^\infty (1+\kappa \re^{-E_n(k, 2)}).
\label{eq:W-Fermi}
\ee
The right hand side can be regarded as the grand partition function of the ideal Fermi-gas, whose energy spectrum
is $E_n(k,2)$. All the information on the partition function is encoded in this spectrum.
The spectral problem in the ABJ(M) Fermi-gas system was studied in \cite{KaMa, Kallen, GHM2} in great detail
(see also \cite{HW, GHM1, WZH, Hatsuda-EQC, HM-Toda, FHM-cluster} for the similar spectral problem associated with topological strings).
The spectrum is completely determined by exact quantization conditions.
The expression \eqref{eq:W-Fermi} guarantees that  $\cW_{\tableau{1}}(\kappa, k, 0)$ is 
an \textit{entire function}
on the complex $\kappa$-plane. It has an infinite number of zeros at $\kappa=-\re^{E_n(k,2)}$ as well as
the trivial one at $\kappa=0$.
It is unclear whether the  function $\cW_R(\kappa, k ,M)$ has such a nice expression.
It would be significant to study the analyticity of $\cW_R(\kappa, k ,M)$ on the $\kappa$-plane.
If $\cW_R(\kappa, k ,M)$ is an entire function, then it is natural to ask whether the zeros of $\cW_R(\kappa, k ,M)$
are also determined by a certain quantization condition or not.

It is also interesting to consider deformations of ABJ theory.
In \cite{MN1, MN2, MN3, HHO}, more general $\cN=4$ superconformal quiver Chern-Simons-matter 
theories were studied in the Fermi-gas approach.
Also, in \cite{Hatsuda-ellip}, the matrix model of ABJM theory on an ellipsoid was studied.
It was found there that this matrix model with a particular value of the deformation parameter
is exactly equivalent to a matrix model that describes the topological string \cite{KM, MZ, KMZ} 
on another Calabi-Yau three-fold, local $\mathbb{P}^2$.
It would be interesting to study Wilson loop VEVs in these deformed theories.

\acknowledgments{
We would like to thank Shinji Hirano, Masazumi Honda, Marcos Mari\~no, Sanefumi Moriyama and Masato Taki 
for useful discussions. 
The work of YH is supported in part by the Fonds National Suisse, subsidies 200021-156995 and by the NCCR 51NF40-141869 
“The Mathematics of Physics” (SwissMAP). The work of KO is supported in part
by JSPS/RFBR bilateral collaboration ``Faces of matrix models in quantum field theory and statistical mechanics''. 
}


\begin{thebibliography}{99}

 \bibitem{ABJM}
 O.~Aharony, O.~Bergman, D.~L.~Jafferis and J.~Maldacena, ``N=6 superconformal Chern-Simons-matter theories, M2-branes and their gravity duals,''
  JHEP {\bf 0810}, 091 (2008)
  [arXiv:0806.1218 [hep-th]].

\bibitem{ABJ} 
  O.~Aharony, O.~Bergman and D.~L.~Jafferis,
  ``Fractional M2-branes,''
  JHEP {\bf 0811}, 043 (2008)
  [arXiv:0807.4924 [hep-th]].

\bibitem{Pestun} 
  V.~Pestun,
  ``Localization of gauge theory on a four-sphere and supersymmetric Wilson loops,''
  Commun.\ Math.\ Phys.\  {\bf 313}, 71 (2012)
  [arXiv:0712.2824 [hep-th]].

 \bibitem{KWY1} 
  A.~Kapustin, B.~Willett and I.~Yaakov,
  ``Exact Results for Wilson Loops in Superconformal Chern-Simons Theories with Matter,''
  JHEP {\bf 1003}, 089 (2010)
  [arXiv:0909.4559 [hep-th]].


\bibitem{Jafferis} 
  D.~L.~Jafferis,
  ``The Exact Superconformal R-Symmetry Extremizes Z,''
  JHEP {\bf 1205}, 159 (2012)
  [arXiv:1012.3210 [hep-th]].

\bibitem{HHL1} 
  N.~Hama, K.~Hosomichi and S.~Lee,
  ``Notes on SUSY Gauge Theories on Three-Sphere,''
  JHEP {\bf 1103}, 127 (2011)
  [arXiv:1012.3512 [hep-th]].

\bibitem{MP-top} 
  M.~Marino and P.~Putrov,
  ``Exact Results in ABJM Theory from Topological Strings,''
  JHEP {\bf 1006}, 011 (2010)
  [arXiv:0912.3074 [hep-th]].

\bibitem{AKMV} 
  M.~Aganagic, A.~Klemm, M.~Marino and C.~Vafa,
  ``Matrix model as a mirror of Chern-Simons theory,''
  JHEP {\bf 0402}, 010 (2004)
  [hep-th/0211098].

\bibitem{DMP1} 
  N.~Drukker, M.~Marino and P.~Putrov,
  ``From weak to strong coupling in ABJM theory,''
  Commun.\ Math.\ Phys.\  {\bf 306}, 511 (2011)
  [arXiv:1007.3837 [hep-th]].


\bibitem{HKPT} 
  C.~P.~Herzog, I.~R.~Klebanov, S.~S.~Pufu and T.~Tesileanu,
  ``Multi-Matrix Models and Tri-Sasaki Einstein Spaces,''
  Phys.\ Rev.\ D {\bf 83}, 046001 (2011)
  [arXiv:1011.5487 [hep-th]].

\bibitem{MP} 
  M.~Marino and P.~Putrov,
  ``ABJM theory as a Fermi gas,''
  J.\ Stat.\ Mech.\  {\bf 1203}, P03001 (2012)
  [arXiv:1110.4066 [hep-th]].

\bibitem{HMO-review} 
  Y.~Hatsuda, S.~Moriyama and K.~Okuyama,
  ``Exact instanton expansion of the ABJM partition function,''
  PTEP {\bf 2015}, no. 11, 11B104 (2015)
  [arXiv:1507.01678 [hep-th]].

\bibitem{HMMO} 
  Y.~Hatsuda, M.~Marino, S.~Moriyama and K.~Okuyama,
  ``Non-perturbative effects and the refined topological string,''
  JHEP {\bf 1409}, 168 (2014)
  [arXiv:1306.1734 [hep-th]].

\bibitem{HMO2} 
  Y.~Hatsuda, S.~Moriyama and K.~Okuyama,
  ``Instanton Effects in ABJM Theory from Fermi Gas Approach,''
  JHEP {\bf 1301}, 158 (2013)
  [arXiv:1211.1251 [hep-th]].

\bibitem{MaMo} 
  S.~Matsumoto and S.~Moriyama,
  ``ABJ Fractional Brane from ABJM Wilson Loop,''
  JHEP {\bf 1403}, 079 (2014)
  [arXiv:1310.8051 [hep-th]].

\bibitem{HoOk} 
  M.~Honda and K.~Okuyama,
  ``Exact results on ABJ theory and the refined topological string,''
  JHEP {\bf 1408}, 148 (2014)
  [arXiv:1405.3653 [hep-th]].

\bibitem{AHS} 
  H.~Awata, S.~Hirano and M.~Shigemori,
  ``The Partition Function of ABJ Theory,''
  PTEP {\bf 2013}, 053B04 (2013)
  [arXiv:1212.2966].

\bibitem{Honda1} 
  M.~Honda,
  ``Direct derivation of "mirror" ABJ partition function,''
  JHEP {\bf 1312}, 046 (2013)
  [arXiv:1310.3126 [hep-th]].
  
\bibitem{DPY} 
  N.~Drukker, J.~Plefka and D.~Young,
  ``Wilson loops in 3-dimensional N=6 supersymmetric Chern-Simons Theory and their string theory duals,''
  JHEP {\bf 0811}, 019 (2008)
  [arXiv:0809.2787 [hep-th]].
  
  \bibitem{CW} 
  B.~Chen and J.~B.~Wu,
  ``Supersymmetric Wilson Loops in N=6 Super Chern-Simons-matter theory,''
  Nucl.\ Phys.\ B {\bf 825}, 38 (2010)
  [arXiv:0809.2863 [hep-th]].
  
  \bibitem{RSY} 
  S.~J.~Rey, T.~Suyama and S.~Yamaguchi,
  ``Wilson Loops in Superconformal Chern-Simons Theory and Fundamental Strings in Anti-de Sitter Supergravity Dual,''
  JHEP {\bf 0903}, 127 (2009)
  [arXiv:0809.3786 [hep-th]].

\bibitem{DT} 
  N.~Drukker and D.~Trancanelli,
  ``A Supermatrix model for N=6 super Chern-Simons-matter theory,''
  JHEP {\bf 1002}, 058 (2010)
  [arXiv:0912.3006 [hep-th]].

\bibitem{KMSS} 
  A.~Klemm, M.~Marino, M.~Schiereck and M.~Soroush,
  ``Aharony-Bergman-Jafferis–Maldacena Wilson loops in the Fermi gas approach,''
  Z.\ Naturforsch.\ A {\bf 68}, 178 (2013)
  [arXiv:1207.0611 [hep-th]].

\bibitem{HHMO} 
  Y.~Hatsuda, M.~Honda, S.~Moriyama and K.~Okuyama,
  ``ABJM Wilson Loops in Arbitrary Representations,''
  JHEP {\bf 1310}, 168 (2013)
  [arXiv:1306.4297 [hep-th]].



  
  
\bibitem{HNS} 
  S.~Hirano, K.~Nii and M.~Shigemori,
  ``ABJ Wilson loops and Seiberg duality,''
  PTEP {\bf 2014}, no. 11, 113B04 (2014)
  [arXiv:1406.4141 [hep-th]].

\bibitem{CGM} 
  S.~Codesido, A.~Grassi and M.~Marino,
  ``Exact results in $ \mathcal{N}=8 $ Chern-Simons-matter theories and quantum geometry,''
  JHEP {\bf 1507}, 011 (2015)
  [arXiv:1409.1799 [hep-th]].


\bibitem{GHM2} 
  A.~Grassi, Y.~Hatsuda and M.~Marino,
  ``Quantization conditions and functional equations in ABJ(M) theories,''
  J.\ Phys.\ A {\bf 49}, no. 11, 115401 (2016)
  [arXiv:1410.7658 [hep-th]].
  

\bibitem{FHM} 
  H.~Fuji, S.~Hirano and S.~Moriyama,
  ``Summing Up All Genus Free Energy of ABJM Matrix Model,''
  JHEP {\bf 1108}, 001 (2011)
  [arXiv:1106.4631 [hep-th]].

\bibitem{Mto}
S. Matsuno and S.~Moriyama, 
``Giambelli Identity in Super Chern-Simons Matrix Model,''
  arXiv:1603.04124 [hep-th].

\bibitem{GKM} 
  A.~Grassi, J.~Kallen and M.~Marino,
  ``The topological open string wavefunction,''
  Commun.\ Math.\ Phys.\  {\bf 338}, no. 2, 533 (2015)
  [arXiv:1304.6097 [hep-th]].


\bibitem{ADKMV} 
  M.~Aganagic, R.~Dijkgraaf, A.~Klemm, M.~Marino and C.~Vafa,
  ``Topological strings and integrable hierarchies,''
  Commun.\ Math.\ Phys.\  {\bf 261}, 451 (2006)
  [hep-th/0312085].


\bibitem{KEK} 
  M.~Hanada, M.~Honda, Y.~Honma, J.~Nishimura, S.~Shiba and Y.~Yoshida,
  ``Numerical studies of the ABJM theory for arbitrary N at arbitrary coupling constant,''
  JHEP {\bf 1205}, 121 (2012)
  [arXiv:1202.5300 [hep-th]].


\bibitem{HO1} 
  Y.~Hatsuda and K.~Okuyama,
  ``Probing non-perturbative effects in M-theory,''
  JHEP {\bf 1410}, 158 (2014)
  [arXiv:1407.3786 [hep-th]].
  

\bibitem{HMO3} 
  Y.~Hatsuda, S.~Moriyama and K.~Okuyama,
  ``Instanton Bound States in ABJM Theory,''
  JHEP {\bf 1305}, 054 (2013)
  [arXiv:1301.5184 [hep-th]].

\bibitem{HMO1} 
  Y.~Hatsuda, S.~Moriyama and K.~Okuyama,
  ``Exact Results on the ABJM Fermi Gas,''
  JHEP {\bf 1210}, 020 (2012)
  [arXiv:1207.4283 [hep-th]].

\bibitem{MS1} 
  S.~Moriyama and T.~Suyama,
  ``Instanton Effects in Orientifold ABJM Theory,''
  arXiv:1511.01660 [hep-th].

\bibitem{Okuyama-ori1} 
  K.~Okuyama,
  ``Probing non-perturbative effects in M-theory on orientifolds,''
  JHEP {\bf 1601}, 054 (2016)
  [arXiv:1511.02635 [hep-th]].

\bibitem{Honda-ori} 
  M.~Honda,
  ``Exact relations between M2-brane theories with and without Orientifolds,''
  arXiv:1512.04335 [hep-th].

\bibitem{Okuyama-ori2} 
  K.~Okuyama,
  ``Orientifolding of the ABJ Fermi gas,''
  JHEP {\bf 1603}, 008 (2016)
  [arXiv:1601.03215 [hep-th]].

\bibitem{MS2} 
  S.~Moriyama and T.~Suyama,
  ``Orthosymplectic Chern-Simons Matrix Model and Chirality Projection,''
  arXiv:1601.03846 [hep-th].

\bibitem{MN-ori} 
  S.~Moriyama and T.~Nosaka,
  ``Orientifold ABJM Matrix Model: Chiral Projections and Worldsheet Instantons,''
  arXiv:1603.00615 [hep-th].

\bibitem{ACDKV} 
  M.~Aganagic, M.~C.~N.~Cheng, R.~Dijkgraaf, D.~Krefl and C.~Vafa,
  ``Quantum Geometry of Refined Topological Strings,''
  JHEP {\bf 1211}, 019 (2012)
  [arXiv:1105.0630 [hep-th]].

\bibitem{AMV} 
  M.~Aganagic, M.~Marino and C.~Vafa,
  ``All loop topological string amplitudes from Chern-Simons theory,''
  Commun.\ Math.\ Phys.\  {\bf 247}, 467 (2004)
  [hep-th/0206164].

\bibitem{Aganagic:2003db} 
  M.~Aganagic, A.~Klemm, M.~Marino and C.~Vafa,
  ``The Topological vertex,''
  Commun.\ Math.\ Phys.\  {\bf 254}, 425 (2005)
  [hep-th/0305132].

\bibitem{KWY2} 
  A.~Kapustin, B.~Willett and I.~Yaakov,
  ``Nonperturbative Tests of Three-Dimensional Dualities,''
  JHEP {\bf 1010}, 013 (2010)
  [arXiv:1003.5694 [hep-th]].
  
\bibitem{Kapustin:2013hpk} 
  A.~Kapustin and B.~Willett,
  ``Wilson loops in supersymmetric Chern-Simons-matter theories and duality,''
  arXiv:1302.2164 [hep-th].

\bibitem{KaMa} 
  J.~Kallen and M.~Marino,
  ``Instanton effects and quantum spectral curves,''
  arXiv:1308.6485 [hep-th].

\bibitem{Kallen} 
  J.~Kallen,
  ``The spectral problem of the ABJ Fermi gas,''
  JHEP {\bf 1510}, 029 (2015)
  [arXiv:1407.0625 [hep-th]].

\bibitem{HW} 
  M.~x.~Huang and X.~f.~Wang,
  ``Topological Strings and Quantum Spectral Problems,''
  JHEP {\bf 1409}, 150 (2014)
  [arXiv:1406.6178 [hep-th]].

\bibitem{GHM1} 
  A.~Grassi, Y.~Hatsuda and M.~Marino,
  ``Topological Strings from Quantum Mechanics,''
  arXiv:1410.3382 [hep-th].

\bibitem{WZH} 
  X.~Wang, G.~Zhang and M.~x.~Huang,
  ``New Exact Quantization Condition for Toric Calabi-Yau Geometries,''
  Phys.\ Rev.\ Lett.\  {\bf 115}, 121601 (2015)
  [arXiv:1505.05360 [hep-th]].

\bibitem{Hatsuda-EQC} 
  Y.~Hatsuda,
  ``Comments on Exact Quantization Conditions and Non-Perturbative Topological Strings,''
  arXiv:1507.04799 [hep-th].

\bibitem{HM-Toda} 
  Y.~Hatsuda and M.~Marino,
  ``Exact quantization conditions for the relativistic Toda lattice,''
  arXiv:1511.02860 [hep-th].

\bibitem{FHM-cluster} 
  S.~Franco, Y.~Hatsuda and M.~Marino,
  ``Exact quantization conditions for cluster integrable systems,''
  arXiv:1512.03061 [hep-th].

\bibitem{MN1} 
  S.~Moriyama and T.~Nosaka,
  ``Partition Functions of Superconformal Chern-Simons Theories from Fermi Gas Approach,''
  JHEP {\bf 1411}, 164 (2014)
  [arXiv:1407.4268 [hep-th]].

\bibitem{MN2} 
  S.~Moriyama and T.~Nosaka,
  ``ABJM membrane instanton from a pole cancellation mechanism,''
  Phys.\ Rev.\ D {\bf 92}, no. 2, 026003 (2015)
  [arXiv:1410.4918 [hep-th]].

\bibitem{MN3} 
  S.~Moriyama and T.~Nosaka,
  ``Exact Instanton Expansion of Superconformal Chern-Simons Theories from Topological Strings,''
  JHEP {\bf 1505}, 022 (2015)
  [arXiv:1412.6243 [hep-th]].

\bibitem{HHO} 
  Y.~Hatsuda, M.~Honda and K.~Okuyama,
  ``Large N non-perturbative effects in $\mathcal{N}=4$ superconformal Chern-Simons theories,''
  JHEP {\bf 1509}, 046 (2015)
  [arXiv:1505.07120 [hep-th]].

\bibitem{Hatsuda-ellip} 
  Y.~Hatsuda,
  ``ABJM on ellipsoid and topological strings,''
  arXiv:1601.02728 [hep-th].

\bibitem{KM} 
  R.~Kashaev and M.~Marino,
  ``Operators from mirror curves and the quantum dilogarithm,''
  arXiv:1501.01014 [hep-th].

\bibitem{MZ} 
  M.~Marino and S.~Zakany,
  ``Matrix models from operators and topological strings,''
  arXiv:1502.02958 [hep-th].

\bibitem{KMZ} 
  R.~Kashaev, M.~Marino and S.~Zakany,
  ``Matrix models from operators and topological strings, 2,''
  arXiv:1505.02243 [hep-th].
  
\end{thebibliography}
\end{document}